\documentclass[pra,twocolumn,amsmath,amssymb,floatfix,superscriptaddress]{revtex4}
\usepackage{graphicx}
\usepackage{bm}

\newcommand{\be}{\begin{equation}}
\newcommand{\ee}{\end{equation}}
\newcommand{\beq}{\begin{eqnarray}}
\newcommand{\eeq}{\end{eqnarray}}

\begin{document}

\title{Microscopic theory of non-adiabatic response in real and imaginary time}

\author{C. De Grandi}
\affiliation{Department of Physics, Yale University, New Haven, Connecticut 06511, USA}

\author{A. Polkovnikov}
\affiliation{Department of Physics, Boston University, Boston, Massachusetts 02215, USA}

\author{A. W. Sandvik}
\affiliation{Department of Physics, Boston University, Boston, Massachusetts 02215, USA}

\date{\today}

\begin{abstract}
We present a general approach to describe slowly driven quantum systems both in real and imaginary time. We highlight many similarities, qualitative  and quantitative, between real and imaginary time evolution. We discuss how the metric tensor and the Berry curvature can be extracted from both real and imaginary time simulations as a response of physical observables. For  quenches ending at or  near the quantum critical point, we show the utility of the scaling theory for detecting the location of the quantum critical point by comparing sweeps  at different velocities. We briefly discuss the universal relaxation to equilibrium of systems after a quench. We finally review recent developments of quantum Monte Carlo methods for studying imaginary-time evolution. We illustrate our findings with explicit calculations using the transverse field Ising model in one dimension.
\end{abstract}

\maketitle

\tableofcontents
\section{Introduction}

Realizing efficient calculations of dynamical properties of interacting quantum systems remains one of the unresolved challenges of modern physics. Even with some recent progress in simulating the dynamics in one dimension using DMRG and related methods~\cite{schollwoeck_05}, as well as exact diagonalization 
(see, e.g., Ref.~[\onlinecite{santos_10}]), most physical systems remain currently out of reach. In a recent work \cite{adi_short} \footnote{We here take the opportunity to point out that in Ref.~[\onlinecite{adi_short}] there is a missing factor of volume $L^d$ in Eq.~(10), which is due to the inverse volume factor 
in the definition of the susceptibility, Eg.~(6) of that work. The correct form should be instead: 
$Q  \approx  v^2 L^d \chi^{(2r+1)}_{\lambda\lambda}$, $F \approx  v^2 L^d \chi^{(2r+2)}_{\lambda\lambda}$.}, we demonstrated that many  difficulties can be overcome by going to imaginary time, where powerful quantum Monte Carlo (QMC) techniques can be used for a wide range of systems. This allows one to study some generic aspects of non-equilibrium dynamics and extract valuable qualitative and quantitative information pertaining also to real-time evolution of interacting systems. We argued that the class of systems which can be analyzed in non-equilibrium setups therefore coincides with those which can be analyzed in equilibrium---those for which the QMC sign problems can be circumvented.

Apart from numerical convenience, imaginary time dynamics also has numerous experimental applications. If we are interested in the dynamics of a subset of 
degrees of freedom of a system, which couple to the rest of the system forming the environment, then the dynamics becomes dissipative. There is no unique 
framework describing dissipative systems, but in many situations one can rely on Langevin dynamics, which is also equivalent to the model A 
dynamics~\cite{halperin_hohenberg}, which is in many cases equivalent to the imaginary time quantum dynamics with extra noise. If the bath has a temperature 
much lower than the driven system then the noise term becomes unimportant. Then the model A dynamics describing the evolution of the real multi-component order parameter 
becomes equivalent to the imaginary time Schr\"odinger equation~\cite{halperin_hohenberg}:
\be
\partial_\tau \psi_j=-\Gamma {\partial F\over \partial\psi_j},
\label{eq1}
\ee
where $\psi_j$ is generally  a multicomponent order parameter, the index $j$ can be either discrete or continuous e.g. corresponding to the spatial coordinate, and $F$ is the free energy of the system. This free energy can explicitly depend on time if we are driving, e.g., external fields which explicitly enter $F$. Such situations were recently considered in Ref.~[\onlinecite{chandran_12}]. Another wide range of applications of imaginary-time quantum dynamics comes from applications to the Kardar-Parisi-Zhang (KPZ) equation and equivalent nonlinear 
Burgers equations which describe the equilibrium behavior of polymers in random media, crystal growth, superconducting flux lines and many other systems (see, e.g., 
Refs.~[\onlinecite{kpz, huse_85}]). For instance, the differential equation describing the partition function of a polymer in a disordered media takes the form 
of the imaginary time Schr\"odinger equation in a random potential, where the role of time is played by the coordinate along the polymer~\cite{huse_85}. Using the 
replica trick, the KPZ equation maps to the imaginary-time Schr\"odinger equation describing bosons with  short-range attractive interactions~\cite{kardar_87}. 
There are several other applications of Eq.~(\ref{eq1}). In this work we will analyze the general properties of the response of  systems described by this
equation together with the real time Schr\"odinger equation in situations where the parameters of the system change slowly in time.

The main purposes of this work are: (i) to further elaborate our earlier findings of Ref.~\cite{adi_short}, (ii) to give a quantitative comparison between 
real and imaginary time evolution for the specific case of the transverse field Ising model, (iii) to discuss how one can extract quantitative information about 
real time correlation functions from the low velocity asymptotics of the imaginary time response. We will also describe how one can extract real and imaginary 
components of the geometric tensor defining the Riemannian metric and the Berry curvature associated with the ground state wave function from both real and 
imaginary time dynamics. We will discuss the application of the nonequilibrium scaling relations for quantum critical systems obtained in 
Refs.~[\onlinecite{deng_08,adi_short, chandran_12, kolodrubetz_11a}]  to accurately locate the quantum critical point and determine the static and dynamic 
critical exponents from the collapse of physical observables. Our results can be useful for: understanding quantum annealing (in particular for finding the 
optimal path in the parameter space), extracting long-time correlation functions and the dynamical exponent for disordered systems, evaluating both real and 
imaginary parts of the geometric tensor for interacting systems, including the Berry curvature and the fidelity susceptibilities. 

The paper is organized as follows. In Sec.~\ref{Kubo} we present the general theory of Kubo response of systems driven with constant velocity both in 
real and imaginary times. We identify the linear and quadratic susceptibilities of the response of arbitrary observables with respect to the velocity using  Adiabatic Perturbation Theory, from here on abbreviated as APT. The 
linear susceptibilities are given by the components of the geometric tensor. In Sec.~\ref{scalingO} we formulate the scaling theory for slowly driven gapless 
systems and systems driven through quantum-critical points and discuss its potential implications for experiments. We also relate the scaling of generic 
observables to the scaling dimension of the relevant components of the geometric tensor. In Sec.~\ref{qmcmethods} we briefly review two complementary Monte-Carlo algorithms~\cite{adi_short, qaqmc} for computing the quantities discussed above. In Sec.~\ref{Ising} we present the exact solution for imaginary-time 
quenches of the transverse-field Ising model in one dimension. From this solution we extract the scaling behaviors for several observables (e.g., excess heat, 
log fidelity, and nearest-neighbour spin-spin correlation functions). We compare these exact results with those obtained using APT and find a very good agreement between them. We also confirm the general scaling relations presented in Sec.~\ref{scalingO}. In Sec.~\ref{imVSreal}  
we  compare the expectation values of different observables in imaginary versus real time quenches. In particular, for diagonal observables like the energy 
and the fidelity, we find a very good qualitative and quantitative agreement between the real and imaginary time evolution. For off-diagonal observables the 
low velocity asymptotics in real and imaginary time are different (given by the imaginary and real components of the geometric tensor) while at high velocities 
they only differ by a numerical factor of the order of one. In Sec.~\ref{detectQCP}  we illustrate how this scaling analysis, either in real or imaginary time, 
can be used to detect the position of the quantum-critical point without the need to vary the system size or the temperature. In Sec.~\ref{relax} we discuss the 
universal behavior of the relaxation of observables (again in both real and imaginary times) to a prethermalized state following a quench near a quantum critical point.

\section{Generalized Kubo response of driven systems}
\label{Kubo}

Originally Kubo response theory was developed to describe transport coefficients for systems in weak electric fields. By now it refers to a general 
linear response theory to a static or time dependent external perturbation~\cite{Mahan}. Let us point that in the Weyl gauge the scalar potential is zero and  the electric field can be thought as the rate of change of the vector potential: $\vec E=1/c\, \partial_t \vec A$. Thus, we can formally view the response to the electric field as the response to the rate of change of the vector potential. Here we extend this analogy and generalize the notion of Kubo response of a system to the rate of change of an arbitrary coupling $\vec{\lambda}$.

Let us consider a system described by a Hamiltonian $\mathcal H[\vec{\lambda}(t)]$, where $t$ is real or imaginary time and $\vec{\lambda}$ is the vector of  coupling constants in the parameter space. To distinguish between real and imaginary time we will reserve the symbol $\tau$ for the latter. In the real time 
case, the dynamics of the system is given by the time dependent Schr\"odinger equation:
\be
i\partial_t \psi(t)=\mathcal H(\vec{\lambda}(t))\psi(t),
\label{sch_eq}
\ee
in imaginary time the dynamics is described by the corresponding equation:
\be
\partial_\tau \psi(\tau)=-\mathcal H(\vec{\lambda}(\tau))\psi(\tau).
\label{sch_eq_ima}
\ee

We will assume that initially the system is prepared at $t=0$ or $\tau=0$ in the ground state. In imaginary time this assumption is not important since, for an evolution lasting sufficiently long time, the initial conditions become irrelevant. In real time our results can be readily generalized to arbitrary stationary 
initial conditions by performing statistical average of the expectation value of an observable over the adiabatically evolved initial density matrix.  We will also assume that the rate of change of the coupling
is sufficiently slow, such that the system remains close to the instantaneous ground state (appropriately rotated with the Berry phase)  at all times during the evolution. We will also make the assumption that the ground state is not degenerate. In this case we can solve the Schr\"odinger equation using the APT. 
The details of the derivation of the first order corrections are presented elsewhere (see Refs.~[\onlinecite{ortiz_2008,degrandi_09}] for real time and 
Ref.~[\onlinecite{adi_short}] for imaginary time). Here we extend the derivations to the second-order terms.

\subsection{Adiabatic perturbation theory in real time}

At the  first order of APT in real time we find that the transition amplitude to the instantaneous state $|n\rangle\neq |0\rangle$ 
is given by~\cite{degrandi_09}:
\be
a_n^{(1)}(t)=-\int_0^t dt_1 \langle n|\partial_{t_1}|0\rangle\exp[-i\Phi_{n0}(t_1,t)],
\label{apt_rt1}
\ee
where $\Phi_{n0}(t_1,t)=\Phi_n(t_1,t)-\Phi_0(t_1,t)$ is the total phase difference accumulated between the ground and the excited states in the time interval $(t_1,t)$,
\be
\Phi_{n}(t_1,t)=\int_{t_1}^t dt_2\left(\mathcal E_n(t_2)-i \langle n|\partial_{t_2}|n\rangle\right),
\label{phase}
\ee
where $\mathcal E_n(t)$ is the energy of the eigenstate $|n\rangle$ at time $t$.
We emphasize that  there is a difference in the limits of integration in Eq.~(\ref{phase}) and Eqs.~(12) and (13) in Ref.~\cite{degrandi_09}. This 
difference is due to additional phase transformation in Eqs.~(6) and (14) in Ref.~[\onlinecite{degrandi_09}]. Going back to the original basis at the end of 
the calculation gives the result above. Note that the overall phase entering in Eq.~(\ref{phase}) can be thought of as a purely dynamical phase coming from 
the gauge invariant energy:
\be
E_n=\mathcal E_n-{v_\alpha}\mathcal A^{(n)}_\alpha,
\ee
where $v_\alpha=\dot\lambda_\alpha$ and $\mathcal A^{(n)}_\alpha=i\langle n|\partial_{\lambda_\alpha}|n\rangle$ is the Berry connection associated with 
the $n$-th energy level. It is easy to see that $E_n$ is invariant under the gauge transformation of the basis states by an arbitrary phase factor:
$|n\rangle\to \exp[i\phi_n(\vec{\lambda}(t))]|n\rangle$. 

Similarly in the second order of  APT one finds
\beq
&& a_n^{(2)}(t)=\sum_{m\neq 0,n} \int_0^t dt_1\int_0^{t_1} dt_2 \langle n|\partial_{t_1}|m\rangle\langle m|\partial_{t_2}|0\rangle\nonumber\\ 
&&~~~~~~~~~~~~~~~~~~~~~~~~~~\times\mathrm e^{-i[\Phi_{nm}(t_1,t)+\Phi_{m0}(t_2,t)]}.
\eeq

Let us point out that the first and second order terms in  APT are not necessarily analytic functions of the adiabatic parameter $\vec v=d_t{\vec \lambda}$. 
Thus, this is a non perturbative expansion. As an example of this, in Sec.~\ref{Ising} we will compare the results obtained within the first-order of APT with 
results from exact diagonalization and find a very good agreement, even when the observables have non-analytic dependence on velocity. In this sense APT does 
not have a formal expansion parameter, it only relies on the fact that the transition probabilities are small. For example, for a Landau-Zener problem the first 
order of APT gives the correct non-analytic dependence of the transition probability on the sweep rate, but leads to a small $\pi^2/9-1$ deviation in the 
prefactor~\cite{degrandi_09}.

For gapped systems, or for gapless systems in sufficiently high dimensions, the leading non-adiabatic corrections to various observables are analytic functions 
of the quench velocity~\cite{polkovnikov_gritsev_08}. Then it is possible to expand the expressions for the transition probabilities as a Taylor series in the velocity. This can be done noting that the integrand in Eq.~(\ref{apt_rt1}) is a product of a slow function (matrix element) and fast function (phase factor) and integrating by parts (see Ref.~\cite{degrandi_09} for details). Keeping terms up to the velocity squared we find that
\begin{widetext}
\be
a_n^{(1)}\approx iv_\alpha {\langle n|\partial_\alpha|0\rangle\over \mathcal E_n-\mathcal E_0}-v_\alpha v_\beta {1\over\mathcal E_n-\mathcal E_0} 
{\partial\over\partial \lambda_\alpha} {\langle n|\partial_\beta|0\rangle\over \mathcal E_n-\mathcal E_0}+iv_\alpha v_\beta 
{\langle n|\partial_\alpha|0\rangle(i\langle n|\partial_\beta|n\rangle-i\langle 0|\partial_\beta|0\rangle)\over (\mathcal E_n-\mathcal E_0)^2},
\label{an1}
\ee
\end{widetext}
where all matrix elements and energies are evaluated at time $t$. It is straightforward to check that if $\mathcal E_n\neq \mathcal E_m$ then
\be
\langle n|\partial_\alpha|m\rangle=-{\langle n|\partial_\alpha H|m\rangle\over \mathcal E_n-\mathcal E_m}.
\label{ident}
\ee
In Eq.~(\ref{an1}) we neglected the additional fast oscillating terms which contain the initial excitations of the system. These terms can be suppressed either 
(i) if the protocol starts smoothly with zero rate or (ii) if the gap in the initial state is very large. The oscillating terms can be further suppressed because 
of various dephasing mechanisms.
 
Similarly to the treatment above we can evaluate the leading order contribution to $a_n^{(2)}$. Again, by neglecting the oscillating terms due to the initial 
excitations in the system, we find that for $n\neq 0$;
\be
a_n^{(2)}\approx -v_\alpha v_\beta \sum_{m\neq 0,n} {\langle n|\partial_\alpha|m\rangle
\langle m|\partial_\beta|0\rangle\over (\mathcal E_n-\mathcal E_0)(\mathcal E_m-\mathcal E_0)}, 
\label{an2}
\ee
and finally the quadratic correction for the ground state amplitude reads:
\be
a_0^{(2)}=-{1\over 2}\sum_{m\neq 0} \left|a_m^{(1)}\right|^2-i v_\alpha v_{\beta}\sum_{m\neq 0}  
{\langle 0|\partial_\alpha|m\rangle\langle m|\partial_\beta|0\rangle\over \mathcal E_m-\mathcal E_0}.
\ee
Combining all the terms up to the second order in $v$ we obtain the result:
\begin{multline}
a_n\approx iv_\alpha {\langle n|\partial_\alpha|0\rangle\over \mathcal E_n-\mathcal E_0}-v_\alpha v_\beta 
{1\over\mathcal E_n-\mathcal E_0} {\partial\over\partial \lambda_\alpha} {\langle n|\partial_\beta|0\rangle\over \mathcal E_n-\mathcal E_0}\\
+v_\alpha v_\beta {\langle n|\partial_\alpha|0\rangle\langle 0|\partial_\beta|0\rangle)\over (\mathcal E_n-\mathcal E_0)^2} -v_\alpha v_\beta \sum_{m\neq 0} 
{\langle n|\partial_\alpha|m\rangle\langle m|\partial_\beta|0\rangle\over (\mathcal E_n-\mathcal E_0)(\mathcal E_m-\mathcal E_0)}. 
\label{an3_real}
\end{multline}

\subsection{Adiabatic perturbation theory in imaginary time}

The APT analysis of the imaginary-time dynamics is very similar to that for the real time case. In Ref.~[\onlinecite{adi_short}] we derived the following 
exact integral equation for the amplitudes of the wave function in the instantaneous basis:
\beq
&&\alpha_n(\tau)=\alpha_n(\tau_f)\nonumber\\
&&+\sum_m \int_{\tau}^{\tau_f} d\tau' \langle n|\partial_{\tau'}|m\rangle \alpha_m(\tau')\mathrm e^{-\int_{\tau'}^{\tau_f} d\tau'' \Delta_{nm}(\tau'')},\phantom{XX}
\label{int_eq_it}
\eeq
where $\Delta_{nm}(\tau)=\mathcal E_n(\tau)-\mathcal E_m(\tau)$ and
\be
\alpha_n(\tau)=a_n(\tau) \exp\left[-\int_{\tau}^{\tau_f} d\tau' \mathcal E_n(\tau')\right].
\label{alpha_n}
\ee
This equation should be supplemented by the normalization condition:
\be
\sum_n |\alpha_n(\tau)|^2=1,
\label{normancond}
\ee
at $\tau=\tau_f$, where $\tau_f$ is the arbitrary final time of interest. From Eq.~(\ref{alpha_n}) it is clear that the coefficient  $\alpha_n(\tau)$ coincides 
with the amplitude $a_n(\tau)$ for the evolved wave function to be in the instantaneous state $|n\rangle$ only at $\tau=\tau_f$. This boundary condition applies 
to the situation where the dynamical process started in a distant past enough for the wave function to become insensitive to the actual initial state (which 
is always possible to satisfy in imaginary time).  Note that, unlike the real time case, the integral equation~(\ref{int_eq_it}) explicitly contains the 
unknown amplitude $\alpha_n(\tau_f)$, which has to be found from the asymptotic boundary condition. As in the real time case, if we deal with the eigenstates 
with a non-zero Berry connection, then one has to use the shifted (complex) energies $E_n=\mathcal E_n+ v_\alpha \langle n |\partial_\alpha|n\rangle=\mathcal E_n- i v_\alpha \mathcal A_\alpha^{(n)}$. Note that the fact that the energies are complex and the "moving Hamiltonian" is non-Hermitean is the consequence of the imaginary time evolution.

From the integral equation (\ref{int_eq_it}) we find that in the leading order of  APT:
\be
\alpha_n^{(1)}(\tau_f)=-\int _{-\infty}^{\tau_f} d\tau \langle n|\partial_\tau|0\rangle 
\mathrm e^{-\int_{\tau}^{\tau_f} d\tau'\Delta_{n0}(\tau')}.
\label{apt_it1}
\ee
For $n\neq 0$ and $\alpha_0^{(1)}=0$, the first-order correction to the ground state amplitude vanishes. Similarly, in the second order of APT we find:
\be
\alpha_n^{(2)}(\tau_f)=-\sum_{m\neq 0,n} \int\limits_{-\infty}^{\tau_f}\!\! d\tau \langle n|\partial_{\tau}|m\rangle \alpha_m^{(1)}(\tau)\mathrm e^{-\int_{\tau}^{\tau_f}\! d\tau' \Delta_{nm}(\tau')}.
\label{apt_it2}
\ee

As in the real-time case one can expand Eqs.~(\ref{apt_it1}) and (\ref{apt_it2}) into a Taylor series in the quench rate:
\begin{widetext}
\be
\alpha_n^{(1)}(\tau_f)\approx -v_{\alpha} {\langle n|\partial_\alpha|0\rangle
\over \mathcal E_n-\mathcal E_0}+v_\alpha v_\beta {1\over \mathcal E_n-\mathcal E_0} {\partial\over\partial
\lambda_\alpha} {\langle n|\partial_\beta|0\rangle\over \mathcal E_n-\mathcal E_0}+v_\alpha v_\beta {\langle n|\partial_\alpha|0\rangle (
\langle n|\partial_\beta|n\rangle-\langle 0|\partial_\beta|0\rangle)\over
(\mathcal E_n-\mathcal E_0)^2},
\label{apt_v1}
\ee
\end{widetext}
where all energies and matrix elements are evaluated at $\tau=\tau_f$. Note that there is a sign difference compared to Ref.~[\onlinecite{adi_short}], 
due to a different definition of the velocity, $v=-\dot\lambda$, used in that work. Similarly, from Eqs.~(\ref{int_eq_it}), (\ref{apt_v1}) and (\ref{apt_it2}) 
we find that for $n\neq 0$:
\be
\alpha_n^{(2)}(\tau_f)\approx v_\alpha v_\beta \sum_{m\neq 0,n} {\langle n|\partial_\alpha|m\rangle\langle m|\partial_\beta|0\rangle\over
(\mathcal E_n-\mathcal E_0)(\mathcal E_m-\mathcal E_0)}.
\label{apt_v2}
\ee

Combining the two expressions above we find that up to $v^2$ terms and for $n\neq 0$:
\begin{multline}
\alpha_n(\tau_f)\approx -v_{\alpha} {\langle n|\partial_\alpha|0\rangle
\over \mathcal E_n-\mathcal E_0}+v_\alpha v_\beta {1\over \mathcal E_n-\mathcal E_0} {\partial\over\partial
\lambda_\alpha} {\langle n|\partial_\beta|0\rangle\over \mathcal E_n-\mathcal E_0}\\
-v_\alpha v_\beta {\langle n|\partial_\alpha|0\rangle\langle 0|\partial_\beta|0\rangle\over
(\mathcal E_n-\mathcal E_0)^2}+v_\alpha v_\beta \sum_{m\neq 0} {\langle n|\partial_\alpha|m\rangle\langle m|\partial_\beta|0\rangle\over
(\mathcal E_n-\mathcal E_0)(\mathcal E_m-\mathcal E_0)}.
\label{an3_im}
\end{multline}
We point that the real time expression for the transition amplitude (\ref{an3_real}) can be formally obtained from the imaginary time expression above by the analytic continuation of the velocity to the complex plane $v\to -iv$. We expect that this will be the case in all orders of expansion in the velocity. However, this continuation might not hold in general when there are additional non-analytic contributions like e.g. an exponential dependence of the transition amplitude on the velocity in the Landau-Zener sweep. There are no analogues of such exponential terms in imaginary time dynamics.  We also point that in order to obtain the complex conjugate of the transition amplitude (\ref{an3_real}) from the imaginary time value (\ref{an3_im}) one needs to analytically continue velocity to positive imaginary axis: $v\to iv$.

The correction to the amplitude for the $n=0$ state  can be found by enforcing the normalization condition (\ref{normancond}) at $\tau=\tau_f$:
\be
\alpha_0^{(2)}(\tau_f)=-{1\over 2}\sum_{m\neq 0} \left|\alpha_m^{(1)}(\tau_f)\right|^2.
\ee

\subsection{Kubo response in the parameter space}
\label{Kubo_res}

Having derived the expressions for the transition amplitudes in real and imaginary time we can next compute the response functions. 
As in Refs.~[\onlinecite{adi_short, asp_berry}], and without loss of generality, we will represent an observable as a \emph{generalized force}:
\be\label{gen_force}
 M_\gamma=-{\langle \psi|\partial_\gamma \mathcal H|\psi\rangle\over \langle \psi|\psi\rangle},
 \ee
 where $\gamma$ is some parameter in the Hamiltonian. The normalization factor in denominator highlights that in imaginary time the wave function should be properly normalized. Clearly  $M_\gamma$ is simply the expectation value of the operator $\mathcal M_\gamma=-\partial_\gamma \mathcal H$.
For example if $\gamma$ is the external magnetic field, then the generalized force $M_\gamma$ is the magnetization, if $\gamma$ is the volume then we have the pressure, if $\gamma$ is the spin-spin interaction then we have 
the spin-spin correlation function, and so on (later on, when presenting the observables for the transverse field Ising model, we will introduce the following generalized forces: excess energy, the log-fidelity and the transverse magnetization, those will be defined according to the definition in Eq.~(\ref{gen_force}) in Sec.~\ref{Ising_obs}).
In the first two orders of the adiabatic perturbation theory (both in real and imaginary time) we find:
\beq\label{M_expect}
M_\gamma\approx M_\gamma^{(0)}&-&\sum_{n\neq 0}\left[ (a_n^{(1)})^\ast \langle n|\partial_\gamma\mathcal H|0\rangle+a_n^{(1)}\langle 0|\partial_\gamma\mathcal H|n\rangle\right]\nonumber\\
&-&\sum_{n}\left[ (a_n^{(2)})^\ast \langle n|\partial_\gamma\mathcal H|0\rangle+a_n^{(2)}\langle 0|\partial_\gamma\mathcal H|n\rangle\right]\nonumber\\
&-&\sum_{n,m\neq 0} (a_n^{(1)})^\ast a_m^{(1)}\langle n|\partial_\gamma\mathcal H|m\rangle,
\eeq
where $M_\gamma^{(0)}=-\langle 0|\partial_\gamma \mathcal H|0\rangle$ is the ground state expectation value. In a more general situation of finite initial 
temperature and real time dynamics, $M_\gamma^{(0)}$ stands for the adiabatic expectation value of $-\partial_\gamma\mathcal H$, i.e., the expectation value with respect to the density 
matrix adiabatically connected to the initial state. Combining this equation with Eqs.~(\ref{an1}) and (\ref{an2}) for the real time case and with (\ref{apt_v1}) 
and (\ref{apt_v2}) for the imaginary case, we find:
-in {\em real time}:
\be
M_\gamma\approx M_\gamma^{(0)}+F_{\gamma\alpha} v_\alpha+[\Pi^{1}_{\gamma\alpha\beta}+\Pi^{2}_{\gamma\alpha\beta}]v_\alpha v_\beta,
\label{kubo_rt}
\ee
- while in {\em imaginary time} we obtain:
\be
M_\gamma\approx M_\gamma^{(0)}-2g_{\gamma\alpha} v_\alpha+[\Pi^{1}_{\gamma\alpha\beta}-\Pi^{2}_{\gamma\alpha\beta}]v_\alpha v_\beta.
\label{kubo_it}
\ee
To the linear order in the velocity, Eq.~(\ref{kubo_rt}) was derived in Refs.~[\onlinecite{avron_11, asp_berry}], and in imaginary time it was derived in 
Ref.~[\onlinecite{adi_short}]. Note again the sign difference in the first term of Eq.~(\ref{kubo_it}) and the result in Ref.~[\onlinecite{adi_short}] 
due to different sign conventions in the definition of $v_\alpha$. In this work $v_\alpha=\partial_\tau \lambda_\alpha$. In the above equations 
$F_{\gamma\alpha}$ and $g_{\gamma\alpha}$ are respectively the Berry curvature and the Riemannian metric tensor, which are related to the imaginary 
(antisymmetric) and real (symmetric) parts of the geometric tensor~\cite{provost_vallee}. Defining the geometric tensor as:
\be
\chi_{\alpha\beta}=\langle 0 |\overleftarrow{\partial_\alpha}\partial_\beta| 0\rangle-\langle 0|\overleftarrow{\partial_\alpha}|0\rangle\langle 0|\partial_\beta|0\rangle,
\label{geom_tens}
\ee
we have:
\beq
F_{\alpha\beta}&=&i\left(\chi_{\alpha\beta}-\chi_{\beta\alpha}\right)=
-2\Im[\chi_{\alpha\beta}], \\
g_{\alpha\beta}&=&{1\over 2}\left(\chi_{\alpha\beta}+\chi_{\beta\alpha}\right)=
\Re[\chi_{\alpha\beta}].
\eeq
The Berry curvature can be also expressed as a curl of the Berry connection~\cite{berry_84}:
\be
F_{\alpha\beta}=\partial_\alpha A_\beta-\partial_\beta A_\alpha,\quad A_\alpha=i\langle 0|\partial_\alpha|0\rangle.
\ee
In the general case of a finite temperature, the expectation values with respect to the ground state in the definition of $F$ and $g$ 
should be substituted by the trace over the density matrix representing the state adiabatically connected to the initial state.

To the second order in the velocity the coefficients of the response functions are defined as follows:
\beq
\Pi^{1}_{\gamma\alpha\beta}&=&\sum_{n,m\neq 0} {\langle 0|\partial_\alpha|n\rangle \langle n|\partial_\gamma \mathcal H|m\rangle\langle m|\partial_\beta|0\rangle\over (\mathcal E_n-\mathcal E_0)(\mathcal E_m-\mathcal E_0)}\nonumber\\
&-&\langle 0|\partial_\gamma \mathcal H|0\rangle \sum_{n\neq 0} {\langle 0|\partial_\alpha|n\rangle\langle n|\partial_\beta|0\rangle\over (\mathcal E_n-\mathcal E_0)^2},\label{Pi_1}\\
\Pi^2_{\gamma\alpha\beta}&=&\sum_{n\neq 0} {\langle 0|\partial_\gamma H|n\rangle\over \mathcal E_n-\mathcal E_0}{\partial\over\partial\lambda_\alpha}{\langle n|\partial_\beta|0\rangle\over
\mathcal E_n-\mathcal E_0}+c.c.+\nonumber\\
&+&\sum_{n,m\neq 0} {\langle 0|\partial_\gamma |n\rangle
\langle n|\partial_\alpha|m\rangle\langle m|\partial_\beta|0\rangle\over
\mathcal E_m-\mathcal E_0}+c.c-\nonumber\\
&-&\sum_{n\neq 0}{\langle 0|\partial_\gamma|n\rangle\langle n|\partial_\alpha|0\rangle\langle 0|\partial_\beta|0\rangle\over \mathcal E_n-\mathcal E_0}+c.c..\label{Pi_2}
\eeq

It is straightforward to see that all the response coefficients $g_{\gamma\alpha}$, $F_{\gamma\alpha}$, $\Pi^1_{\gamma\alpha\beta}$ and $\Pi^2_{\gamma\alpha\beta}$ 
are gauge invariant, i.e. invariant under arbitrary basis transformations: 
\[
|m(\vec{\lambda})\rangle\to \mathrm e^{i f_m (\vec{\lambda})}|m\rangle.
\]

There is a special class of observables in this regard: those that commute with the Hamiltonian in the final state. Examples of these observables 
include the Hamiltonian itself (i.e., the energy of the system) as well as its various moments; the expectation value of any other conserved quantity; 
the probability to remain in the ground state, which is known as the fidelity, or in any particular eigenstate of the final Hamiltonian; 
the diagonal entropy of the system, and others. For these diagonal observables the linear response term in Eqs.~(\ref{kubo_rt}) and (\ref{kubo_it}) vanishes, as
so does the response coefficient $\Pi^2_{\gamma\alpha\beta}$. Therefore, the leading-order non-adiabatic response is given by the following quadratic 
response function, which is the same both for real and imaginary time dynamics:
\be
M_\gamma\approx M_\gamma^{(0)}+\Pi^1_{\gamma\alpha\beta} v_\alpha v_\beta,
\ee
\be
\Pi^{1}_{\gamma\alpha\beta}=\sum_{n\neq 0} {\langle 0|\partial_\alpha|n\rangle \langle n|\partial_\beta|0\rangle\over (\mathcal E_n-\mathcal E_0)^2}
\left[(\partial_\gamma \mathcal H)_{nn}-(\partial_\gamma \mathcal H)_{00}\right],
\ee
where $(\partial_\gamma \mathcal H)_{nn}\equiv \langle n|\partial_\gamma \mathcal H|n\rangle$.

\subsection{Relation to the zero-frequency limit of the response functions}

The components of the geometric tensor~\cite{venuti_07, adi_short} and the second-order susceptibility $\Pi^{1,2}_{\gamma\alpha\beta}$ can 
be expressed through the non-equal time correlation functions of the generalized forces. Using Eq.~(\ref{ident}), the geometric tensor (\ref{geom_tens}) for 
a non-degenerate ground state can be rewritten as: 
\be
\chi_{\alpha\beta}=\sum_{n\neq 0} {\langle 0|\partial_\alpha H |n\rangle\langle n|\partial_\beta H|0\rangle\over (\mathcal E_n-\mathcal E_0)^2}.
\label{geom_tens1}
\ee
Next we recall the identity:
\be
{1\over (\mathcal E_n-\mathcal E_0)^2}=-\lim_{\epsilon\to 0+}\int_0^\infty d\xi\, \xi\, \mathrm e^{-\epsilon \xi-i (\mathcal E_n-\mathcal E_0)\xi},
\ee
with which one can rewrite the geometric tensor (\ref{geom_tens1}) as an integral from the retarded correlation function:
\be
\chi_{\alpha\beta}=-\int_{0}^\infty d\xi\, \xi \langle 0|\partial_\alpha \mathcal H(\xi)\partial_\beta \mathcal H(0)|0\rangle_c \mathrm e^{-\epsilon  \xi},
\label{geom_tens2}
\ee
where the subindex $c$ implies connected and
\be
\partial_\alpha \mathcal H(\xi)=\mathrm e^{i\mathcal H \xi}\partial_\alpha \mathcal H \mathrm e^{-i\mathcal H \xi} 
\ee
is the Heisenberg representation of the generalized force $\partial_\alpha \mathcal H$ at the point of measurement. Note that here $\xi$ is an auxiliary variable, which 
is not related to the time evolution during the dynamical process. One can also rewrite Eq.~(\ref{geom_tens2}) in the imaginary time Heisenberg representation by formally 
rotating to Euclidean time: $\xi\to -i\tau$. 

One can also rewrite Eq.~(\ref{geom_tens2}) through a derivative of the Fourier transform of the non-equal time correlation function:
\be
\chi_{\alpha\beta}=-i\partial_\omega G_{\alpha\beta}(\omega)\bigl|_{\omega=0},
\ee
where
\be
G_{\alpha\beta}(\omega)=\int_0^\infty dt \mathrm e^{i\omega t} \langle 0|\partial_\alpha \mathcal H(t)\partial_\beta\mathcal H(0)|0\rangle_c\, \mathrm e^{-\epsilon t}.
\ee
From this expression we see that the metric tensor and the Berry curvature are given by the imaginary and real parts of the frequency derivative of the 
corresponding correlation function:
\be
g_{\alpha\beta}=\partial_\omega \Im G_{\alpha\beta}(\omega)\bigr|_{\omega=0},\; F_{\alpha\beta}=2\partial_\omega\Re G_{\alpha\beta}(\omega)\bigr|_{\omega=0}.
\ee

The components of the quadratic susceptibility $\Pi^{1,2}_{\gamma\alpha\beta}$ can be represented through the time-time correlation functions in a similar fashion. For example:
\beq
\Pi^{1}_{\gamma\alpha\beta}&&=-\int_0^\infty dt_1\int_0^\infty dt_2\;\mathrm e^{-\epsilon (t_1+t_2)}\, t_1t_2\nonumber\\
&& \bigl[\langle 0|\partial_\alpha\mathcal H(t_1)\partial_\gamma \mathcal H\partial_\beta\mathcal H(t_2)|0\rangle\nonumber\\
&&-\langle 0|\partial_\alpha\mathcal H(t_1)\partial_\beta\mathcal H(t_2)|0\rangle\langle 0|\partial_\gamma \mathcal H|0\rangle\bigr].
\eeq

\subsection{Equivalent derivation of the Kubo response via a generalized Galilean transformation}

The linear Kubo response given by Eqs. (\ref{kubo_rt}) and (\ref{kubo_it}) can be derived in a simple and intuitive way by going to a moving frame. 
The time dependent Schr\"odinger equation (\ref{sch_eq}) can be rewritten in a comoving basis of the instantaneous Hamiltonian as
\be
i\tilde\partial_t|\psi\rangle =\left(H-v_\alpha P_\alpha\right) |\psi\rangle=H_{\rm eff}|\psi\rangle,
\label{galilean}
\ee 
where the time derivative $\tilde \partial_t$ acts only on the coefficients of the expansion of the wave function in the comoving basis, while $P_\alpha$ 
is a generalized momentum operator with respect to the parameter $\lambda_\alpha$. It can be formally defined through the matrix elements in the instantaneous basis;
\be
\langle n|P_\alpha|m\rangle=i\langle n|\partial_\alpha|m\rangle=-i {\langle n|\partial_\alpha H|m\rangle\over \mathcal E_n-\mathcal E_m}.
\label{P_alpha}
\ee

It is easy to see that $P_\alpha$ is a Hermitian operator. This follows e.g. from differentiating the identity
\be
\langle n| m\rangle=\delta_{mn}
\ee
with respect to $\alpha$. Alternatively one can note that the states $|m(\vec\lambda)\rangle$ can be obtained from some fixed basis corresponding to e.g. $\vec\lambda_0$ by some unitary transformation:
\be
|m(\vec\lambda)\rangle = U_{mn} |n(\vec{\lambda_0}\rangle.
\ee
For a non-degenerate spectrum this unitary operator corresponds to the adiabatic evolution of the Hamiltonian. Then in the moving frame the Hamiltonian in the Schr\"odinger equation will clearly acquire an extra correction
\be
-i U^{-1} \partial_t U=-i v_\alpha U^{-1}\partial_\alpha U=-v_\alpha P_\alpha.
\ee
Clearly the momentum operator $P_\alpha=i U^{-1}\partial_\alpha U$ is the same as above in Eq.~(\ref{P_alpha}).

The RHS of Eq.~(\ref{galilean}) extends the conventional Galilean transformation of the Hamiltonian in the moving frame. Indeed let us assume that we have a system of $N$ 
interacting particles in an external potential, which depends on time via:
\[
V(\vec x_1,\dots \vec x_N,t)=V(\vec x_1-\vec X(t),\dots \vec x_N-\vec X(t)),
\]
where $\vec X(t)$ is a time dependent vector defining the moving frame where the potential is stationary. This vector $\vec X(t)$ can also denote a center of 
mass coordinate of an interacting system. Then Eq.~(\ref{galilean}) is indeed the Galilean transformation where: 
\[
\vec v=\dot{\vec X},\; {\rm and}\; \vec P=\sum_{j=1}^N\vec p_j,
\]
are the usual velocity and  momentum operator. 

\begin{figure}[ht]
\includegraphics[width=8.5cm, clip]{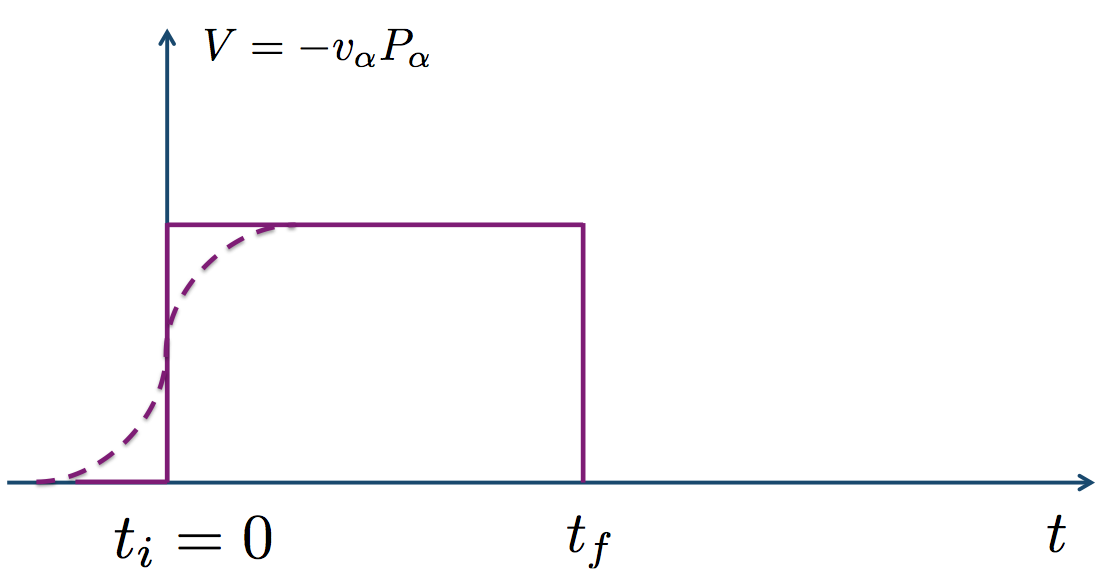}
\caption{(Color online) Schematic representation of the perturbation in the effective Hamiltonian in the moving frame for the case of constant velocity 
[see Eq.~(\ref{galilean})]. At the initial time $t_i=0$ the velocity instantaneously changes from zero to a finite value and at the point of measurement 
$t=t_f$ the velocity effectively drops back to zero reflecting the transformation back to the lab frame. The shape of the pulse can be smoothen at the 
initial time by turning on the perturbation slowly (dashed line). }
\label{fig:pulse_v}
\end{figure}

Close to the adiabatic limit the additional term in the effective Hamiltonian in Eq.~(\ref{galilean}) can be treated as a perturbation. A very similar equation 
can be written in imaginary time with the (non-Hermitean) Hamiltonian in the moving frame: $H_{\rm eff}=H -i v_\alpha P_{\alpha}$. At the initial moment of time we turn on the velocity 
and at the moment of measurement we effectively turn it off. Indeed, we may view the result of the instantaneous measurement as a result of a sudden quench of the 
velocity going back to zero, which is equivalent to going back to the original lab frame. Thus, for a simple protocol where the velocity suddenly turns on at time 
$t_i=0$ and the measurement is done at time $t_f$ our perturbation in the effective Hamiltonian looks like a pulse (see Fig.~\ref{fig:pulse_v}).
Now the results of the APT, e.g., Eq.~(\ref{an1}), leading to the Kubo formulas in real and imaginary time are easily understood as those of the ordinary 
perturbation theory with respect to the velocity dependent term in the effective Hamiltonian; see Eq.~(\ref{galilean}). Indeed, if the time of the pulse is short, 
then we can apply an ordinary time dependent perturbation theory where the transition amplitude is proportional to the matrix element of the perturbation 
integrated over time, i.e.: 
\be
\int_0^t v_\alpha P_\alpha dt\approx \delta \lambda_\alpha P_\alpha.
\ee 
Thus, in this limit of a short pulse we simply reproduce the ordinary perturbation theory where the transition amplitude is proportional to the change 
of the coupling constant. Conversely, in the long time limit, the response is very different: to leading order of perturbation theory the system is excited 
at the initial time, then it freely evolves in a moving basis, after which it is excited again by the quench at $t=t_f$. Thus, the transition amplitude is the 
sum of two terms, corresponding to the initial and final quenches. The contribution due to the initial quench comes in as a rapidly changing Rabi phase. As we 
discussed above, this phase averages to zero due to any dephasing mechanism, or it can be simply suppressed by a slow turning-on of the velocity. If this is the 
case, then we can reproduce the transition amplitudes~(\ref{an1}) and (\ref{apt_v1}) by using an ordinary static perturbation theory with respect to the 
perturbation $V=v_\alpha P_\alpha$ for real time protocols and $V=iv_\alpha P_\alpha$ for imaginary time protocols.

\subsection{Extracting the Berry curvature from the imaginary time dynamics. Analytic continuation of the imaginary time dynamical response to real time.}
\label{sec:berry_it}

Equations~(\ref{kubo_rt}) and (\ref{kubo_it}) show that one can extract the Berry curvature as a linear response to the quench velocity in real time and the metric 
tensor from the imaginary time response. Real time dynamics of course corresponds to physical processes and thus can be directly realized in experiments. However, 
with the exception of one dimensional systems, there are very limited numerical tools which would allow for real time simulations. Imaginary-time dynamics has 
the advantage that it is amenable to powerful QMC as we first showed with the non-equilibrium QMC (NEQMC) approach in Ref.~\cite{adi_short} and later with a 
different quasi-adiabatic QMC (QAQMC) scheme in Ref.~\cite{qaqmc}. We will briefly review these approaches in Sec.~\ref{qmcmethods}. 

Because of the opportunities offered by the NEQMC and QAQMC methods, it would be very practical to have a way to extract the Berry curvature using imaginary-time 
dynamics as well. This indeed becomes possible if we evolve left and right wave functions in imaginary time with different velocities, $v^L_\alpha$ and $v^R_\alpha$, 
and evaluate the following expectation value:
\be
\tilde M_\beta=i\left[\langle \psi_L(\tau)| \partial_\beta H |\psi_R(\tau)\rangle-\langle \psi_R(\tau)| \partial_\beta H |\psi_L(\tau)\rangle\right],
\ee
where $\psi_L(\tau)$ and $\psi_R(\tau)$ are the solutions of the imaginary time Schr\"odinger equation for the  two protocols characterized by different velocities. 
It is important that both $\psi_L(\tau)$ and $\psi_R(\tau)$ are evaluated at the same point in the parameter space. Then it is easy to see that in the linear order 
in the velocities $v_L$ and $v_R$ we have:
\be
\tilde M_\beta\approx \mathcal F_{\beta\alpha} (v^R_\alpha-v^L_\alpha).
\label{berry_it}
\ee
In particular, if $v^L_\alpha\to 0$, i.e. if the left wave functions is the instantaneous ground state, then the equation above is equivalent to the real time 
linear response~(\ref{kubo_rt}). Very similar results apply to real-time simulations, i.e., if we compute $\tilde M_\beta$ in a real-time protocol we will get the 
linear response proportional to the real part of the metric tensor.

The result (\ref{berry_it}) opens the possibility of extracting the Berry curvature from  numerical simulations in imaginary time. Note that the Berry curvature 
is non-zero only if the time reversal symmetry is broken, and, hence, the wave function is complex. In these situations QMC simulations usually suffer from the sign 
problem. However, it is important to mention that the time reversal symmetry can be broken by the parameter $\beta$, which does not enter the time evolution and 
only appears in the definition of the observable we measure. In these situations, the wave function always remains real and the sign problem may be avoidable. 

In imaginary time dynamics left and right states corresponding to opposite velocities naturally occur for the asymmetric expectation values. For a closely related QAQMC algorithm this issue was discussed in detail in Ref.~\cite{qaqmc}. Let us define the protocol where the coupling $\lambda$ changes in imaginary time in the symmetric fashion in the interval: $[0,2T]$: $\lambda(\tau)=\lambda(2T-\tau)$.  Then denoting the imaginary time evolution operator
\be
U(\tau_1,\tau_2)=T_\tau \exp\left[-\int_{\tau_1}^{\tau_2} \mathcal H(\tau) d\tau\right]
\ee
we can write a generally asymmetric value of arbitrary observable $M_\gamma (\tau,2T-\tau)$ as
\be
M_\gamma(\tau,2T-\tau)={\langle \psi_0 | U(2T,\tau) \mathcal M_\gamma U(\tau,0) |\psi_0\rangle\over 
\langle \psi_0|U(2T,0)|\psi_0\rangle},
\label{m_asym1}
\ee
where $\psi_0$ is the initial state and $\tau\in[0,2T]$. Such expectation values are very straightforward to evaluate for any $\tau$ in e.g. Monte-Carlo simulations (see also discussion below following Eq.~(\ref{atdef})). It is easy to see that in the middle point of the evolution $\tau=T$ this expectation value becomes equivalent to the result of the imaginary time dynamics discussed in the previous sections. Away from the symmetric point $\tau\neq T$ the left and right states effectively evolve with the opposite velocities so the expectation value above effectively becomes 
\be
M_\gamma(\tau, 2T-\tau)\bigg|_{\tau\neq T}\approx {\langle \psi_L(-v)| \mathcal M_\gamma |\psi_R(v)\rangle\over \langle \psi_L(-v)| \psi_R(v)\rangle}.
\label{m_asym}
\ee
As it is discussed in Ref.~\cite{qaqmc} the interval around $\tau=T$ where the crossover between symmetric and asymmetric asymptotics happens vanishes as $v\to 0$.

From the analytic properties of the wave-function discussed below Eq.~(\ref{an3_im}) it is clear that the real time expectation value of the observable $M_\gamma$  can be obtained by the analytic continuation of Eq.~(\ref{m_asym}) to the imaginary velocity $v\to -iv$. Indeed as we noted earlier to get the real time result for the wave function and its complex conjugate one needs to analytically continue the velocity to the imaginary axis in the opposite ways $v\to \pm iv$. However, because in Eq.~(\ref{m_asym}) $\psi_L$ and $\psi_R$ are evaluated at opposite velocities both have to be analytically continued in the same way $v\to -iv$. As a result the analytic continuation works for the observable $M_\gamma$. It is interesting that formally this analytic continuation is valid perturbatively to all orders in $v$. In particular, this implies that the leading asymptotic of off-diagonal observables in the asymmetric points in (\ref{m_asym1}) will be given by the Berry curvature multiplied by the velocity and for the diagonal observables the leading asymptotic will be quadratic in velocity but will have a negative sign (i.e. opposite in sign to the real time asymptotic).
 
\subsection{Extension to  non-linear quench protocols}

The linear response analysis carried out above was based on the assumption that near the point of measurement the quench velocity $v_\alpha$ is non-zero. This 
is the case for generic protocols. However, it is also possible to especially design protocols where one approaches the point of measurement $t=t_f$ with some other 
power law characterized by the exponent $r\geq 0$ (see also Ref.~\onlinecite{degrandi_10}):
\be
\vec\lambda(t)\approx \vec \lambda_f+\vec v_r {(t_f-t)^r\over r!}.
\label{lambda_rt}
\ee
Here the parameter:
\be
\vec v_r=(-1)^r d^r \vec\lambda(t)/dt^r|_{t=t_f}
\ee
is the adiabaticity parameter, which plays the role of the quench amplitude for sudden quenches 
($r=0$), quench velocity for linear quenches ($r=1$), quench acceleration for quadratic quenches ($r=2$), and so on. We should note that, with the definition 
(\ref{lambda_rt}) for a linear quench, $\vec v_r=-\dot{\vec\lambda}$ is actually a negative velocity. However, the convention above is more natural for non-linear
protocols, in particular for noninteger values of $r$. In Ref.~\cite{chandran_12} it was also analyzed the case when $r$ is negative and  were found different 
behaviors. In this work we will not be concerned with such situations. 

To evaluate the leading non-adiabatic response for quenches according to (\ref{lambda_rt}) with arbitrary $r$ we need to perform the asymptotic analysis of 
Eqs.~(\ref{apt_rt1}) and (\ref{apt_it1}) using Eq.~(\ref{lambda_rt}) for the dependence of $\vec\lambda(t)$ and taking the asymptotic limit $|\vec v_r|\to 0$. 
In this limit the matrix elements and the energy spectrum entering Eqs.~(\ref{apt_rt1}) and (\ref{apt_it1}) are approximately constant. Then, as an example,
in imaginary time Eq.~(\ref{apt_it1}) reduces to:
\be
\alpha_n^{(1)}=-\langle n|\partial_\alpha|0\rangle\int^{\vec\lambda_f} d\lambda_\alpha \mathrm e^{-\Delta_{n0}(\tau_f-\tau(\vec\lambda))}, 
\ee
where the matrix element $\langle n|\partial_\alpha|0\rangle$ and the energy difference $\Delta_{n0}$ are computed at the point of measurement: 
$\vec\lambda=\vec\lambda_f$. The integral is taken along the actual path $\vec\lambda(\tau)$ and is readily evaluated in the long-time limit:
\be
\alpha_n^{(1)}\approx v_{r,\alpha} {\langle n|\partial_\alpha|0\rangle\over (\mathcal E_n-\mathcal E_0)^r}.
\ee
For the linear quench this expression reduces to the leading term in Eq~(\ref{apt_v1}), keeping in mind that for $r=1$ we have $\vec v_{r}=-\vec v$;
see Eq.~(\ref{lambda_rt}). Similarly, in real time we find
\be
a_n^{(1)}\approx \mathrm (-i)^r v_{r,\alpha} {\langle n|\partial_\alpha|0\rangle\over (\mathcal E_n-\mathcal E_0)^r}.
\ee
Using these excitation amplitudes we can easily find linear response expressions for the generalized forces extending 
Eqs.~(\ref{kubo_rt}) and (\ref{kubo_it}). 

In {\em real time}:
\be
M_\gamma\approx M_\gamma^{(0)}+\left[(-i)^r\chi_{\gamma\alpha}^{r+1}+(i)^r\chi_{\alpha\gamma}^{r+1}\right] v_\alpha+O(v^2) ,
\label{kubo_rt1}
\ee
while in {\em imaginary time}:
\be
M_\gamma\approx M_\gamma^{(0)}+\left[\chi_{\gamma\alpha}^{r+1}+\chi_{\alpha\gamma}^{r+1}\right] v_\alpha+O(v^2).
\label{kubo_it1}
\ee
Here we have defined:
\beq
\chi^{r+1}_{\gamma_\alpha}&=&\sum_{n\neq 0} 
{\langle 0|\partial_\gamma \mathcal H|n\rangle\langle n|\partial_\alpha \mathcal H|0\rangle\over (\mathcal E_n-\mathcal E_0)^{r+1}}\nonumber\\
&=& i^{r+1}\int dt\, {t^r\over r!}\mathrm e^{-\epsilon t} \langle 0|\partial_\gamma \mathcal H(t)\partial_\alpha\mathcal H(0)|0\rangle.
\eeq

As for the linear quenches the difference between real and imaginary time transition amplitudes is contained in the phase factors. One can always eliminate these 
factors artificially by considering separately left and right states, e.g., in the imaginary time evolution and forming the appropriate linear combinations 
similarly to Sec.~\ref{sec:berry_it}. For diagonal observables such as the energy or generic observables which are measured not instantaneously after 
the quench at $t=t_f$ but after allowing the system to relax to the diagonal ensemble, the phases in Eqs.~(\ref{kubo_rt1}) and (\ref{kubo_it1}) do not matter. 
Thus, to leading order in $v$ the responses in  real and imaginary time coincide and are given by:
\be
M_\gamma\equiv \langle \psi|\mathcal M_\gamma|\psi\rangle\approx M_\gamma^{(0)}+\Pi^{1, r}_{\gamma\alpha\beta}v_\alpha v_\beta,
\ee
where:
\be
\Pi^{1,r}_{\gamma\alpha\beta}=\sum_{n\neq 0} 
{\langle 0|\partial_\alpha\mathcal H|n\rangle\langle n|\partial_\beta\mathcal H|0\rangle\over (\mathcal E_n-\mathcal E_0 )^{2r+2}} 
\left[\langle n|\mathcal M_\gamma|n\rangle-\langle 0|\mathcal M_\gamma|0\rangle\right].
\ee
This expression opens a way of measuring the symmetric part of the geometric tensor in real-time experiments. For example, one could measure the excess heat:
$\mathcal M_\gamma=\mathcal H$ for a protocol with $r=1/2$. In this case $\Pi^{1,1/2}_{\gamma\alpha\beta}$ reduces to the metric tensor $g_{\alpha\beta}$.

Furthermore, we note that  if the power of the quench is $r=4$, the  equations  (\ref{kubo_rt1})  and (\ref{kubo_it1}) become identical.
This suggests the interesting fact that by probing a system through a \emph{quartic} quench ($\sim t^4$), we get the same
response in real and imaginary time dynamics.

\section{Universal scaling near quantum critical points}
\label{scalingO}

The linear response theory presented above allows one to associate deviations from adiabaticity of various observables through different susceptibilities. 
These susceptibilities are in turn expressed through integrals of non-equal time correlation functions and, in particular, are very sensitive to their long 
time asymptotics. However, in gapless regimes, specifically, near continuous phase transitions, these susceptibilities may diverge and the linear response 
theory then breaks down. In these situations one can extend standard scaling theory of continuous phase transitions to non-equilibrium setups. 

The scaling hypothesis originally introduced by Pokrovsky-Patashinsky and Kadanoff (see, e.g., Refs.~[\onlinecite{zinn-justin,sachdev}] for references) is based 
on the conjecture that universal physics near continuous phase transitions depends on the microscopic parameters only through the correlation length. For quantum 
phase transitions, which are relevant for our work here, the correlation length universally diverges with the tuning parameter~\cite{sachdev} as:
\be
\xi_\lambda(\vec\lambda)\sim {1\over |\vec\lambda-\vec\lambda_c|^\nu},
\ee
where $\nu$ is the correlation length exponent. For a multi-dimensional parameter space one can have different exponents $\nu$ along different directions. 
In this case there is more than one correlation length. Then, in finite size systems the scaling hypothesis states that the expectation value of any observable 
can be written in the following scaling form:
\be
M_\gamma={\rm const}+L^{-\mu_\gamma} f_\gamma(L/\xi_\lambda(\vec\lambda)),
\label{scaling_eq}
\ee
where the first constant term represents a non-universal non-critical contribution to $M_\gamma$, while $\mu_\gamma$ is a universal number which defines the scaling dimension 
of the operator $\mathcal M_\gamma$, $L$ is the system size and $f_\gamma(x)$ is a universal function. Various two-point correlation functions can be represented in a 
similar form~\cite{sachdev} where instead of the system size we use the separation between  points. By definition the scaling dimension of the tuning parameter is $1/\nu$.

In general, the scaling dimensions of the coupling $\lambda_\alpha$ and the corresponding generalized force $\mathcal M_\alpha$ are independent. But there is an 
important exception, namely, when the coupling $\lambda_\alpha$ is relevant, i.e, this coupling drives the system to or away from the critical point. Then the scaling 
dimension of the product $(\vec\lambda-\vec\lambda_c)_\alpha \mathcal M_\alpha$ must be equal to $z$, the scaling dimension of the energy, and thus we must have 
$\Delta_\alpha+1/\nu=z$. If we are dealing with marginal perturbations which keep the system in the gapless regime, e.g. which lead to renormalization of the 
velocity, then we generally have to set $\nu\to\infty$ (corresponding to scaling dimension of $\lambda_\alpha$ equal to zero) leading to $\Delta_\alpha=z$. In 
this work we will assume that the driving term in the Hamiltonian is either relevant or marginal but that the observable $\mathcal M_\gamma$ has an arbitrary 
scaling dimension $\mu_\gamma$. The generalization to situations where the driving term is irrelevant is straightforward. 

As it was argued in Refs.~[\onlinecite{deng_08, degrandi_10, adi_short, kolodrubetz_11,kolodrubetz_11a, chandran_12, qaqmc}] the scaling ansatz (\ref{scaling_eq}) can be 
extended to non-equilibrium situations if we add the quench velocity as another scaling variable. In particular, since $v_\alpha=d\lambda_\alpha/dt$ we can expect that 
the scaling dimension of the velocity is ${\rm dim}[v_\alpha]={\rm dim}[\lambda_\alpha]-{\rm dim}[t]=\nu^{-1}+z$. This scaling dimension implies that there should be 
another length scale associated with the velocity $\vec v$:
\be
\xi_v\sim {1\over |\vec v|^{\nu\over z\nu+1}}.
\ee
For generic power law protocols characterized by the exponent $r$ [see Eq.~(\ref{lambda_rt})] the above result immediately generalizes to:
\[
\xi_v\sim v^{-{\nu\over z\nu r+1}}.
\] 
This length scale is indeed the Kibble-Zurek length, which was first introduced by Zurek~\cite{zurek_85} for classical phase transitions in relation to the Kibble-Zurek mechanism~\cite{Kibble1976, zurek_85} and later reintroduced in  Refs.~[\onlinecite{ap_adiabatic, zurek_05}] for quantum phase transitions. Physically $\xi_v$ describes the length scale beyond which the system effectively freezes 
and cannot follow the instantaneous ground state.  For multicritical points with more than one direction corresponding to different critical exponents, $\xi_v$ can depend 
on the direction of the quench~\cite{Mukherjee2011}. Now one can use this additional length scale to extend the ansatz for both real and imaginary time evolution (\ref{scaling_eq}) to:
\be
M_\gamma(\vec \lambda,\vec v)={\rm const}+L^{-\mu_\gamma} f_\gamma(L/\xi_\lambda(\vec\lambda), L/\xi_v).
\label{scaling_neq}
\ee
Here $\vec\lambda$ and $\vec v$ describe the coupling and its rate of change at the point where we perform the measurement of our observable $\mathcal M_\gamma$.

The asymptotics of the scaling function $f_\gamma$ can be often determined from qualitative considerations. Thus, if $\xi_v\gg \xi_\lambda$, then the system effectively 
behaves adiabatically and we should recover the scaling behavior pertaining to the static equilibrium. In the opposite limit, depending on the ratio of $\xi_v$ and $L$, we should recover  a similar crossover between the linear response discussed earlier and non-equilibrium universal scaling. For $\xi_v\gg L$ we have $f_\gamma(L/\xi_v)\sim (L/\xi_v)^{z+1/\nu}\propto v$ if there is a non-vanishing relevant component of the geometric tensor or $f_\gamma(L/\xi_v)\sim (L/\xi_v)^{2z+2/\nu}\propto v^2$ in cases where we expect quadratic scaling with the  velocity, e.g., if $M_\gamma$ is a diagonal observable. This asymptotic predicts a non-trivial scaling of the observables with the system size for very slow protocols. In the opposite limit $L\gg \xi_v$, we expect the extensive observables to scale linearly with  the system 
size, i.e., $f_\gamma(L/\xi_v)\sim (L/\xi_v)^{\mu_\gamma+1}$. While for intensive observables  $M_\gamma$ should saturate to a constant value independent on the system 
size, such that $f_\gamma (L/\xi_v)\sim (L/\xi_v)^{\mu_\gamma}$. 

These simple considerations well reproduce the scaling behaviors derived earlier. For example, we expect that the density of defects is intensive and that the number 
of defects has scaling dimension zero (more accurately this statement applies to the log fidelity~\cite{degrandi_10}) and  thus:
\be
n\sim {1\over L^d} f(L/\xi_v)\sim 1/\xi_v^d\sim v^{d\nu/(z\nu+1)},
\ee
is the well known Kibble-Zurek scaling form~\cite{zurek_85}. Similarly we can recover the scaling of the excess energy density for quenches ending near a 
quantum-critical point: $Q\sim v^{(d+z)/(z\nu+1)}$, which follows from noting that the scaling dimension of the Hamiltonian is $z$ (or, equivalently, that the scaling 
dimension of the Hamiltonian density is $d+z$). These scaling considerations equally apply to imaginary-time (dissipative) and real-time dynamics. The only difference 
is that in the low-velocity limit $\xi_v\ll L$ the response is given by different susceptibilities, which, however, have the same scaling properties. Note that while
in this paper we focus on quantum-critical points, the general considerations equally apply to thermal transitions, as it was emphasized in Ref.~[\onlinecite{chandran_12}].

\section{Time-evolving quantum Monte Carlo algorithms}
\label{qmcmethods}

One of the primary reasons for considering imaginary-time dynamics is that it is amenable to numerical simulation with modified
QMC methods. This way, one can go beyond one dimension (where DMRG is applicable in real time) for a rather broad class of systems for 
which sign problems can be avoided. This class coincides with that for which equilibrium QMC methods can be applied. 

Standard QMC algorithms can be classified into {\it finite-temperature} methods, where the goal is to compute a quantum-mechanical
thermal average of the form:
\begin{equation}
\langle A\rangle = \frac{1}{Z}{\bf Tr}\{ {\rm e}^{-\beta H}\},~~~~
Z = {\bf Tr}\{ {\rm e}^{-\beta H}\},
\label{qmcz}
\end{equation}
and {\it ground-state projector} methods, where some operator $P(\beta)$ is applied to a ``trial state'' $|\Psi_0\rangle$,
such that $|\Psi_\beta\rangle = P(\beta)|\Psi_0\rangle$ approaches the ground state (up to an irrelevant normalization factor) 
when $\beta \to \infty$ and an expectation value (which is what normally is computed, although one can also stochastically generate 
the wave function)
\begin{equation}
\langle A\rangle = \frac{1}{Z}\langle \Psi_\beta|A|\Psi_\beta\rangle,~~~~ Z = \langle \Psi_\beta|\Psi_\beta\rangle,
\end{equation}
approaches the true ground state expectation value, $\langle A\rangle \to \langle 0| A|0\rangle$. For the projector, one can use 
$P(\beta)={\rm e}^{-\beta \mathcal{H}}$ with large $\beta$ or a high power of the Hamiltonian, $P(m)=\mathcal{H}^m$. If in the latter case one uses 
$m \propto N\beta$ with $\beta$ of the former approach, the same rate of convergence applies for a given system volume $N$ (which follows, e.g., from a 
series expansion of the exponential, which for large $\beta$ is dominated by powers of the order $n=\beta |E_0|$, where $E_0$ is the ground 
state energy). 

The differences between $T>0$ and $T=0$ projector methods can be thought of in terms of different boundary conditions in imaginary time: The trace taken 
at $T>0$ in (\ref{qmcz}) corresponds to periodic boundaries while the projector methods correspond to opening up these boundaries and replacing them with the
ones corresponding to the trial state. Completely open boundary conditions correspond to the trial state being the equal superposition of all states
in the basis used. 

The time-evolving QMC methods we have developed are essentially modified projector algorithms. In the original NEQMC approach the exponential operator
${\rm e}^{-\beta \mathcal{H}}$ for a fixed Hamiltonian is replaced by the Schr\"odinger evolution operator in imaginary time,
\begin{equation}
U(\tau)=T_\tau {\rm exp}\left [ - \int_{\tau_0}^\tau d \tau' \mathcal{H}[\lambda(\tau')] \right ],
\label{utau}
\end{equation}
where $T_\tau$ indicates time ordering. As in equilibrium QMC schemes, there are several ways to deal with the exponential. 
In the context of spins and bosons, the most frequently used methods
are based on (i) the Suzuki-Trotter-decomposition, which leads to world-line methods \cite{hirsch82}, (ii) the continuous-time version of world-lines
(e.g., the {\it worm algorithm} \cite{prokofev98}), and (iii) the Taylor expansion leading to the SSE method \cite{sandvik91,sandvik92} (see Ref.~\cite{sandvik10} 
for a recent review of these approaches). The latter two methods are not affected by any approximations (beyond statistical sampling errors), while (i) has a
discretization error. 

In the NEQMC algorithm a series expansion is employed for (\ref{utau}) in the non-equilibrium case, while in the more recently introduced QAQMC method the 
power $\mathcal{H}^m$ of the Hamiltonian used in standard projector methods is replaced by a product of evolving Hamiltonians. It was shown in Ref.~\cite{qaqmc} 
that the product evolution reproduces imaginary-time Schr\"odinger dynamics up to the leading corrections in $v$ to the adiabatic evolution. In practice, 
this kind of method is easier to implement than the NEQMC scheme, and, moreover, one can obtain results for all times between the initial and final Hamiltonian 
in a single run. We here briefly review the two methods.

\subsection{Non-equilibrium Schr\"odinger approach}
\label{alg1}

In the NEQMC scheme first proposed in \cite{adi_short} the exponential in (\ref{utau}) is expanded in a power-series and applied to an initial state
$|\Psi(0)\rangle$:
\begin{eqnarray}
&&|\Psi(\tau)\rangle = \sum_{n=0}^\infty \int_{\tau_0}^\tau d\tau_n \int_{\tau_0}^{\tau_n} d\tau_{n-1} \cdots \int_{\tau_0}^{\tau_3} d\tau_2 \times  \nonumber \\
&&~~~ \int_{\tau_0}^{\tau_2} d\tau_1  [\mathcal{-H}(\tau_n)] \cdots [\mathcal{-H}(\tau_1)]|\Psi(0)\rangle.
\label{qmc1}
\end{eqnarray}
The Hamiltonian is a sum of terms,
\be
\mathcal{H} = -\sum_{i=1}^{N_{\rm op}} H_i,
\ee
where the index $i$ can refer to lattice sites, links, etc., and $N_{\rm op}$ is the total number of these operators. A minus sign has been included for convenience.
The operator product in (\ref{qmc1}) is then written as a sum over all strings of the operators $H_i$. Truncating at some maximum power $n=M$ 
(adapted to cause no detectable error---see \cite{sandvik10} for a discussion of this issue in the SSE method) and introducing a trivial unit 
operator $H_{0}=1$, one obtains:
\begin{eqnarray}
&&|\Psi(\tau)\rangle =  \sum_{H} \frac{(M-n)!}{(\tau-\tau_0)^{M-n}} \int_{\tau_0}^\tau d\tau_m \cdots \int_{\tau_0}^{\tau_3} d\tau_2  \times \nonumber \\
&&~~~ \int_{\tau_0}^{\tau_2} d\tau_1  H_{{i_m}}(\tau_m) \cdots H_{{i_2}}(\tau_2) H_{{i_1}}(\tau_1)|\Psi(0)\rangle,
\label{qmc2}
\end{eqnarray}
where $i_p \in \{0,\ldots, M\}$, $\sum_{H}$ denotes the sum over all possible sequences of $M$ operators $H_i$ and $n$ is the number of indices 
$i_p \not =0$. 

At this stage a basis $\{|\alpha\rangle\}$ should be chosen. For spin systems, this would normally be the standard basis of the $z$ spin components:
$|\alpha\rangle = |S^z_1,\ldots,S^z_N\rangle$.  If the initial state $|\Psi(0)\rangle$ has some simple known expansion in this basis, $|\Psi(0)\rangle =        
\sum_\alpha c_\alpha |\alpha\rangle$, this can be used in (\ref{qmc2}). The scheme is particularly simple when using the equal superposition, e.g.,
$|\Psi(0)\rangle = \prod_i (\uparrow_i + \downarrow_i)$ for an $S=1/2$ system, but other states can be used as well. For models with spin-isotropic 
interactions, such as the Heisenberg model, it is easy to use amplitude-product states in the singlet sector \cite{sandvik10b}. One can also start 
with the ground state of some Hamiltonian $\mathcal{H}(\lambda_0)$, by adding to (\ref{qmc1}) a projection with that fixed Hamiltonian before the 
quench with the time dependent $\mathcal{H}[\lambda(\tau)]$ is applied.

For practical implementations of the QMC scheme,
the terms $H_i$ of $\mathcal{H}$ should have the property that $H_i|\alpha\rangle = h_i(\alpha)|\alpha'\rangle$, where $|\alpha'\rangle$ is a
basis state. In the standard spin basis, this implies that $H_i$ is either a diagonal or off-diagonal operator (i.e., $i$ denotes not only a lattice 
unit but also refers to either a diagonal or off-diagonal part). A string of operators and their associated time values, along with a state $|\alpha\rangle$ 
then constitute a configuration, and the QMC simulation amounts to importance-sampling of these configurations, which strongly resemble those of a
path integral.

To guarantee the absence of a sign problem we need to place certain conditions (which are not always possible to satisfy) on the matrix elements $h_i(\alpha)$, 
i.e., the product of all matrix elements corresponding to a term in (\ref{qmc2}) has to be positive. While this is a limitation of the QMC approach in 
general, the class of accessible models is still large and includes highly non-trivial and important systems (see Ref.~\cite{annurev} for a recent review
of quantum spin models without sign problems).

Expectation values $\langle \Psi(\tau)|A|\Psi(\tau)\rangle/\langle \Psi(\tau)|\Psi(\tau)\rangle$ are computed by sampling the normalization
$\langle \Psi(\tau)|\Psi(\tau)\rangle$ written with (\ref{qmc2}). The procedures are very similar to those used in the standard SSE and ground-state
projector methods \cite{sandvik10,sandvik03,sandvik10b}. The main difference is that each operator is associated with a time value. The simplest 
way to deal with the times is to sample them completely independently of the operator and state updates, i.e., the latter are performed with fixed time values, 
according to one of the standard schemes \cite{sandvik10,sandvik03,sandvik10b}, and the times are updated without changes in the operators and states. 
A segment of $m$ times, $\tau_i,\ldots,\tau_{i+m-1}$, can be simultaneously updated by generating $m$ numbers within the range $(\tau_{i-1},\tau_{i+m})$,
then order these times according to a standard scheme scaling as $\log(m)$ \cite{numerical07}. The ordered set is then inserted in place of the old set
of times, with a Metropolis acceptance probability obtained from (\ref{qmc2}). The number $m$ can be adjusted to give an acceptance probability 
close to $1/2$. 

As discussed in Sec.~\ref{sec:berry_it}, in addition to conventional expectation values, it is also useful to study asymmetric expectation values defined 
with two different time evolution operators $U$ and $V$, e.g., corresponding to different velocities: $\langle \psi_0 | V^* A  U|\psi_0 \rangle/\langle \psi_0 
| V^* U|\psi_0 \rangle$. This can be done with a simple generalization of the above NEQMC algorithm.

\subsection{Quasi-adiabatic approach}
\label{alg2}

One may ask whether the role of the time integrals in Eq.~(\ref{qmc1}) is crucial. Clearly, they are needed in order to obtain a mathematically exact
expansion of the time evolution operator (\ref{utau}), but one can, in fact, also formulate a scheme similar to time evolution without these integrals,
by acting on the initial ground state $|\Psi_0\rangle$ of $\mathcal{H}[\lambda_0]$ with a product of $M$ evolving Hamiltonians:
\begin{equation}
P_{M,1}=[-\mathcal{H}(\lambda_M)]....[-\mathcal{H}(\lambda_2)][-\mathcal{H}(\lambda_1)],
\label{p1m}
\end{equation}
where we consider the parameter changing according to
\begin{equation}
\lambda_t= \lambda_0+t\Delta_\lambda,
\label{lambdat}
\end{equation}
and $\Delta_\lambda=[\lambda_{t+1}-\lambda_t]/M$ is the single-step change in the tuning parameter. One can also consider a non-linear grid of points,
but here we focus on the linear evolution. It is clear that $|\psi_M\rangle=P_{M,1}|\Psi_0\rangle$ approaches the ground state $|\Psi_M\rangle$
of the final Hamiltonian $\mathcal{H}[\lambda_M]$ in the limit of large $M$ (up to an irrelevant normalization). 

In Ref.~\cite{qaqmc} it was also 
demonstrated, using APT, that $N\Delta_\lambda$, where $N$ is the system volume, plays the role of a velocity $v$, and that $|\psi_M\rangle$ captures 
the leading non-adiabatic corrections in $v$ to the imaginary-time Schr\"odinger evolution. This is sufficient for recovering all the dynamical 
susceptibilities that we discussed in this work. 

Moreover, one can also consider generalized (asymmetric) expectation values of the form:
\begin{equation}
\langle A\rangle_t = \frac{\langle \Psi(\lambda_0)| P_{1,M} P_{M,t+1}AP_{t,1}|\Psi(\lambda_0)\rangle}{\langle \Psi(\lambda_0)| P_{1,M} P_{M,1}|\Psi(\lambda_0)\rangle},
\label{atdef}
\end{equation}
where only the special case $t=M$ corresponds to a true quantum mechanical expectation value but also the generic $t\not=M$ quantities contain useful
dynamic information and obey dynamic finite-size scaling. A significant advantage of QAQMC over the NEQMC approach is then that one can compute $\langle A\rangle_t$
for all $t$ simultaneously in a single simulation for operators $A$ that are diagonal in the basis used. Such a simulation amounts to generating terms (paths)
contributing to the normalization $\langle \Psi(\lambda_0)| P_{1,M} P_{M,1}|\Psi(\lambda_0)\rangle$ and successively measuring diagonal observables after propagation 
of the state with $t$ operators, for $t$ on a suitable grid.

\begin{figure}
\includegraphics[width=8.4cm, clip]{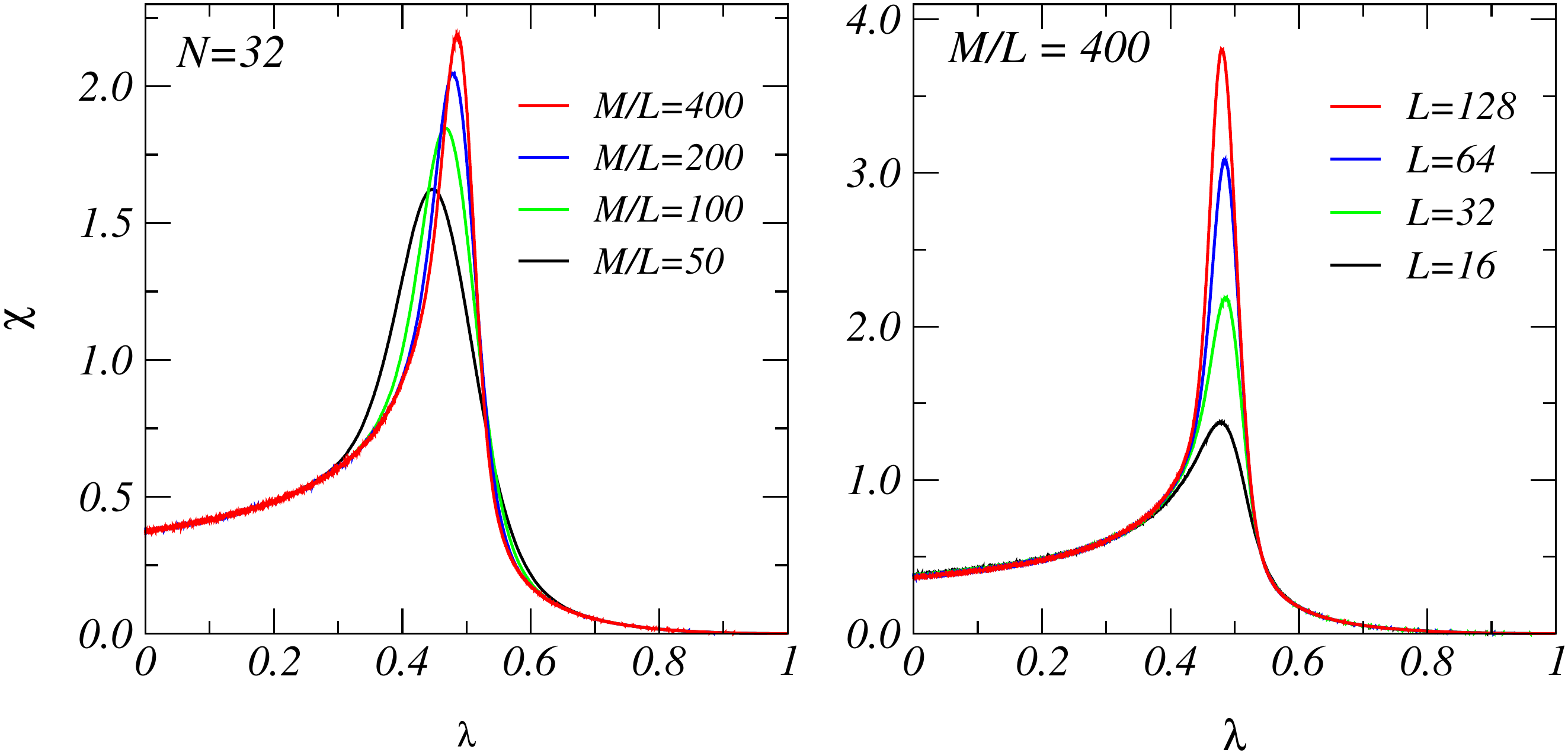}
\caption{\small (Color online) Magnetization fluctuation, Eq.~(\ref{susmz}), of the 1D transverse-field Ising model (see Eq.~(\ref{isA})) in QAQMC calculations with
different length $M$ of the operator string for an $L=32$ system (left) and for different system sizes at fixed $M$ (right). The Ising and field terms
are $J=\lambda$, $g=1-\lambda$, so that the quantum-critical point is at $\lambda=1/2$. The whole $0 \le \lambda \le 1$ curve was obtained in a single simulation.}
\label{susqaqmc}
\end{figure}

Figure \ref{susqaqmc} shows examples of results obtained with the QAQMC method in simulations of the 1D transverse-field Ising model, which we introduce in the next section. The quantity shown is the
magnetization fluctuation,
\begin{equation}
\chi = N\left ( \left\langle m_z^2 \right\rangle -   \left\langle |m_z| \right\rangle ^2 \right),
\label{susmz}
\end{equation}
which exhibits a peak close to the known quantum-critical point at $g=J$. The peak grows both as a function of the size $L$ and $m$, and one can subject
the data to various forms of finite-size and finite-velocity scaling, examples of which are discussed in Ref.~\cite{qaqmc}.

As an alternative to computing expectation value based on evolving the same state from the left and the right in (\ref{atdef}), one can also carry out QAQMC
simulations as a {\it one-way evolution}. In the simplest case, the left state $\langle \psi_L|$ is the ground state of $\lambda_0$  and the right
state  $\langle \psi_R|$ is the ground state of $\lambda_M$. The single sequence (\ref{p1m}) between these state will then smoothly connect them, and, again,
this evolution captures the leading non-adiabatic corrections in $v$ to the standard Schr\"odinger dynamics, with $v \propto L\Delta_\lambda$. As we discussed earlier (see Sec.~\ref{sec:berry_it}) asymmetric expectation values including one way evolution can be used for computing the Berry curvature in the system.

\section{Transverse field Ising model in one dimension} \label{Ising}

We now investigate the results of the previous sections by analyzing  quenches in real or imaginary time of the one-dimensional (1D) transverse field Ising model. 
This model maps onto free fermions and, thus, it is easily solvable. It was used to rigorously demonstrate the universal scaling relations both for equilibrium phase 
transitions~\cite{sachdev} and various aspects of quantum dynamics, including the Kibble-Zurek scaling~\cite{dziarmaga_05}. Furthermore, using a closely related model it was recently demonstrated that 
the universality of slow quantum dynamics does not rely on integrability~\cite{kolodrubetz_11a}. We also point out that this model was extensively used to study the dynamics following sudden quenches~(see e.g. Refs.~\cite{McCoy1971,calabrese2011,Igloi2011,Schuricht2012}). Here we will use this model again for the purpose of a detailed comparisons 
between real and imaginary time dynamics, to establish that the universal aspects are identical. We will also demonstrate how one can use this universality to accurately 
extract the equilibrium transition point and the critical exponents using non-equilibrium protocols. 


The 1D transverse-field Ising model is defined by the Hamiltonian:
\be
\mathcal H=-g\sum_j \sigma_j^x-J\sum_{\langle ij\rangle}\sigma_i^z\sigma_{j}^z,
\label{isA}
\ee
where $\sigma^x$ and $\sigma^z$ are Pauli matrices and $\langle ij\rangle$ are nearest neighbours sites. The parameters $g$ and $J$ control the nature of 
the quantum state: for $g/J > 1$ the system is a \emph{quantum paramagnet}, while for $g/J < 1$ it is in a  \emph{magnetically ordered} phase. The point
$(g/J)_c=1$ of this spin chain is a quantum critical point (QCP) separating the two different phases.
 
\subsection{Imaginary time quench of the transverse field Ising model:  exact solution}
\label{exact}

We investigate  the dynamical response of the system in the vicinity of the QCP  by changing in time either $g$ or $J$. Quenches in real time of the Ising model 
have been considered previously~\cite{degrandi_09},  while less is known in imaginary time. In Ref.~[\onlinecite{adi_short}] we have fixed $g=1$ and considered 
a particular quench protocol for $J$: $J(\tau)=1+v\tau$, starting in the ground state at $\tau_0=-1/v$ and ending at the QCP at $\tau=0$. We can implement a 
similar process fixing $J=1$ and changing $g$ as $g(\tau)=1-v\tau$. If instead we change $g$ as $g(\tau)=1+v\tau$, then we approach the QCP from the initially 
ordered (ferromagnetic) phase. All these protocols give a very similar scaling behavior of the observables. Thus, for extracting analytical results we will 
focus on the particular protocol:
 \be\label{gprot}
  g(\tau)=1+\lambda(\tau), \qquad \lambda=-v \tau.
 \ee
In the next section we will illustrate our results with numerical simulations, in which we also consider different protocols.
  
\subsubsection{Spectrum of the Ising chain}

 It is well known that the Hamiltonian~(\ref{isA}) can be mapped to that one of non-interacting fermions using the Jordan-Wigner 
transformation~\cite{sachdev, dziarmaga_05}. Because of  translational invariance, the relevant excited states are only those which contain pairs 
of quasi-particles with opposite momenta. As a result, in this reduced Hilbert space the Hamiltonian of the system splits into a direct sum of 
Hamiltonians describing two-level systems with the states: $\uparrow\rangle_k$ and $|\downarrow\rangle_k$ corresponding to empty and filled fermionic 
levels with momenta $(k,-k)$, respectively:
\be
\mathcal H=\sum_{k>0} \mathcal H_k,
\ee
where:
\be\label{Hk}
\mathcal H_k= -2[g -\cos(k)] \hat{\sigma}^z +2 \sin(k) \hat{\sigma}^x.
\ee
Each of these Hamiltonians has the following eigenvectors:
\be
|+\rangle_k=\left(\begin{array}{c} \sin(\theta_k/2) \\ -\cos(\theta_k/2)\end{array}
\right),\quad |-\rangle_k=\left(\begin{array}{c} \cos(\theta_k/2) \\
\sin(\theta_k/2)\end{array}\right),
\label{pm_k}
\ee
where:
\be
\tan \theta_k=\frac{\sin(k)}{\cos(k)-g},
\label{tan_thet}
\ee
corresponding to the eigenenergies $E_k^{\pm}=\pm \varepsilon_k$ with $\varepsilon_k= 2\sqrt{1+g^2-2g\cos(k)}$.  It is easy to check that:
\be\label{mat_elem}
_k\langle +\vert \partial_g|-\rangle_k=-{1\over 2} {\sin(k)\over 1+g^2-2g\cos(k)},
\ee
where we have used $\partial_g={\partial\theta_k\over \partial g} \partial_{\theta_k}$.

For a linear quench protocol $g(\tau)=1+\lambda=1-v\tau$, the imaginary time Schr\"odinger equation (\ref{sch_eq_ima}) splits into 
a sum of independent differential equations:
\beq \label{eq:Full1}
&&\dot a_k=2 [1-v\tau -\cos(k)] a_k- 2\sin(k) b_k,\\
&&\dot b_k=-2 \sin(k)a_k-2 [1-v\tau -\cos(k)] b_k.\label{eq:Full2}
\eeq
In the limit of $\tau \to -\infty$ we have $\theta_k \to 0$ and the eigenstates (\ref{pm_k}) simply become:
\be
|+\rangle_k^{\tau \to -\infty}=\left(\begin{array}{c} 0 \\ 1 \end{array} \right),\quad |-\rangle_k^{\tau \to -\infty}=\left(\begin{array}{c} 1 \\ 0 \end{array} \right). 
\ee
At the critical point $g_c=1$, corresponding to $\tau=0$, we have $\theta_k=-\left( \frac{\pi -k}{2} \right)$ so that:
\be\label{Eig:qcp}
|+\rangle_k^{\tau=0}=\left(\begin{array}{c} \sin(\frac{\pi -k}{4}) \\ \cos(\frac{\pi -k}{4})\end{array}
\right),\quad |-\rangle_k^{\tau=0}=\left(\begin{array}{c} \cos(\frac{\pi -k}{4}) \\
-\sin(\frac{\pi -k}{4})\end{array}\right).
\ee

\subsubsection{Linearized spectrum: exact solution}

In the adiabatic limit, where only low momentum modes contribute to the excitations, we  work with the linearized spectrum of the Ising model. 
Then the Hamiltonian  (\ref{Hk}) simplifies to $\mathcal H_k= 2 v\tau  \hat{\sigma}^z +2 k \hat{\sigma}^x$ and Eqs.~(\ref{eq:Full1}) and (\ref{eq:Full2}) reduce to:
\beq \label{lz1}
&&\dot a_k=-2 v\tau b_k- 2 k b_k,\\
&&\dot b_k=-2 k a_k+2 v\tau b_k.\label{lz2}
\eeq
These equations can be solved exactly analytically (we point out that this problem is the imaginary-time counterpart of the \emph{half} Landau Zener 
(LZ) problem analyzed in~Refs.~\onlinecite{vitanov_99, damski_06, degrandi_09}). It is convenient to  rescale the variables, $\tau\to\tau/\sqrt{v}$, 
$k\to q\sqrt{v}$, and differentiate both of these equations with respect to time. We then obtain $\ddot{a}_q-\left(4 \tau^2+4 q^2 -2\right)a_q=0 $ 
and $ \ddot{b}_q-\left(4 \tau^2+4 q^2 +2\right)b_q=0$. Each of these equations is of the type $\ddot y -(4 x^2+c) y=0$, which has two generic solutions:
$y_1(x)=e^{-x^2}  {_1 F_1}\left(\frac{ c}{8}+\frac{1}{4},\frac{1}{2},2 x^2\right)$ and  
$y_2(x)=(-1)^{-1/4}(2 x) e^{-x^2} { _1 F_1}\left(\frac{ c}{8}+\frac{3}{4},\frac{3}{2},2 x^2\right)$,
where $_1 F_1(a,b,z)$ is the confluent hypergeometric function. The generic solutions for Eqs.~(\ref{lz1}) and (\ref{lz2})  are therefore:
\begin{widetext}
\be\label{LZsolut2}
a_q(\tau)=c_1 e^{-\tau^2}
 {_1 F_1}\left(\frac{q^2}{2},\frac{1}{2},2 \tau^2\right)+c_2 (-1)^{-1/4}(2 \tau)  e^{-\tau^2}
 {_1 F_1}\left(\frac{q^2}{2} +\frac{1}{2},\frac{3}{2},2 \tau^2\right),
\ee
\be\label{LZsolut1}
b_q(\tau)=c_3 e^{-\tau^2}
 {_1 F_1}\left(\frac{q^2}{2} +\frac{1}{2},\frac{1}{2},2 \tau^2\right)+c_4 (-1)^{-1/4}(2 \tau)  e^{-\tau^2}
 {_1 F_1}\left(\frac{q^2}{2} +1,\frac{3}{2},2 \tau^2\right).
\ee
\end{widetext}
The coefficients $c_1,c_2,c_3,c_4$ are determined by the initial conditions on the wave function at time $\tau_0$ and two auxiliary conditions, e.g., 
$\dot{a_q}\rvert_{\tau=0}=-2q\, b_k(0) $ and $\dot{b_q}\rvert_{\tau=0}=-2q\, a_q(0)$ (continuity at $\tau=0$). These last two equations give 
$c_2=(-1)^{5/4}q \, c_3$, $c_4=(-1)^{5/4}q \,c_1$, while $c_1$ and $c_3$ are set by the requirement that the system was in its ground state in the 
distant past: $a_q(\tau_0\to -\infty)=1$ and  $b_q(\tau_0\to -\infty)=0$.

Using the expansion of the hypergeometric function when $|z|\to \infty$:
\begin{eqnarray}
 {_1 F_1}(a,b,z) = e^z z^{a-b} \frac{\Gamma(b)}{\Gamma(a)}+ (-z)^{-a} \frac{\Gamma(b)}{\Gamma(b-a)}+\mathcal{O}\left(\frac{1}{z}\right)\nonumber
\end{eqnarray}
we find that:
\be
c_3=-\frac{q}{\sqrt{2}} \frac{\Gamma(q^2/2+1/2)}{\Gamma(q^2/2+1)}c_1.
\ee
The resulting probability of being in the excited state at the end of the evolution for $\tau=0$ is found by overlapping the final wave function 
with the excited state at $\tau=0$:
 \be
\vert + \rangle^{\tau=0}=\frac{1}{\sqrt{2}}\left(\begin{array}{c} 1 \\ 1 \end{array} \right).
\ee
Therefore we have:
\begin{eqnarray}
p^{LZ}_{\rm ex}(q)& =&\frac{1}{2}\frac{|a_q(0)+b_q(0)|^2}{|a_q(0)|^2+|b_q(0)|^2}\nonumber \\
& =& \frac{1}{2}\frac{\left(1-\frac{q}{\sqrt{2}} \frac{\Gamma(q^2/2+1/2)}{\Gamma(q^2/2+1)}\right)^2}{1+\left(\frac{q}{\sqrt{2}} 
\frac{\Gamma(q^2/2+1/2)}{\Gamma(q^2/2+1)}\right)^2},
\end{eqnarray}
where we point out that, since ${_1 F_1}(a,b,0)=1$ for any $a$ and $b$, we have $a_q(0)=c_1$ and $b_q(0)=c_3$. Restoring the dependence on $v$ 
using the substitution $q \to k/\sqrt{v}$ we obtain:
\begin{eqnarray}\label{pLZ}
p^{LZ}_{\rm ex}(k,v)& = \frac{1}{2}\frac{\left[\Gamma\left(\frac{k^2}{2 v}+1\right)-\frac{k}{\sqrt{2 v}}\Gamma\left(\frac{k^2}{2 v}+\frac{1}{2}\right)\right]^2}{\Gamma\left(\frac{k^2}{2 v}+1\right)^2+\frac{k^2}{2 v}\Gamma\left(\frac{k^2}{2 v}+\frac{1}{2}\right)^2}.
\end{eqnarray}

\begin{table}
\begin{center}
\begin{tabular}{|c|c|c|c|}
\hline
~Observable~ & ~$vL^2\ll 1$~ & ~$vL^2\gg 1$~ & ~$vL^2\gg 1$~ \tabularnewline
& Exact sol. and APT & Exact sol. & APT \tabularnewline
\hline
~$M_x$~ & $[{1}/{16}] v L^2$
& $0.26\sqrt{v}L$ & $0.296\sqrt{v}L$
\tabularnewline
\hline
$Q$ & $[{7\zeta(3)}/{128}\pi^3] v^2 L^3$ & $0.0265\, vL$
& $0.0273\, vL$ \tabularnewline
\hline
$F$ & $[{1}/{6144}] v^2 L^4 $ & $0.0276 \sqrt{v} L$ & $0.0314\sqrt{v} L$\tabularnewline
\hline
\end{tabular}
\hfill{}
\vskip-1mm
\caption{Scaling of several observables in the 1D transverse field Ising model after a linear quench in imaginary time [Eq.~(\ref{gprot})], with quench  velocity $v$ and system size $L$. The second and third column show the asymptotics of the exact solutions for: the finite-size adiabatic limit, $vL^2\ll 1$, and thermodynamic 
adiabatic limit $vL^2\gg 1$. The last column shows the scaling results within APT for $vL^2\gg 1$. For $vL^2\ll 1$ the perturbative (APT) and exact expressions are identical.}
\label{table1}
\end{center}
\vskip-3mm
\end{table}

Furthermore we note that to correctly define the final amplitudes $\alpha_{+,-}(\tau=0)$ on the $|+\rangle$ and $|-\rangle$ eigenstate, we need to properly normalized the coefficients in Eqs.~(\ref{LZsolut2})  and~(\ref{LZsolut1}) to satisfy the condition  in Eq.~(\ref{normancond}).

\subsection{Observables}
\label{Ising_obs}
In the following we present some observables that describe the response of the system to the quench in Eq.~(\ref{gprot}). Their scaling behavior is derived 
from the exact solution using the excitation probability in Eq.~(\ref{pLZ}). The results are summarized in Table \ref{table1}, where they are also compared 
with the correspondent scaling found by adiabatic perturbation theory (APT). 

\subsubsection{Excess energy Q}

We consider the total excess energy of the system (energy above the instantaneous ground state energy):
\be
Q=\langle \mathcal{H} \rangle-\langle \mathcal{H} \rangle_0.
\ee
The scaling predictions in Eq.~(\ref{kubo_it}) apply to this case when the observable $M_\gamma\equiv \langle \psi|\mathcal M_\gamma|\psi\rangle$ is associated 
with the Hamiltonian operator: $\mathcal M_\gamma=H$, according to our definition of generalized force in Eq.~(\ref{gen_force}).
In this case, since we are dealing with a diagonal operator, the geometric tensor is identically zero [see Eqs.~(\ref{geom_tens}) and (\ref{geom_tens1})] and the 
response is quadratic and, in this case of a quench of a single parameter, is proportional to a single component $\Pi^{1}_{E\lambda\lambda}$ (Eq.~\ref{Pi_1}):
\be
\Pi^1_{E\lambda\lambda}=\sum_{n\neq 0}{|\langle 0|\partial_\lambda H|n\rangle|^2\over (\mathcal E_n-\mathcal E_0)^3}.
\ee
The response to the quench, both in real and imaginary time, is:
\be\label{scal_heat}
Q\approx v^2 \Pi^1_{E\lambda\lambda}.
\ee
From the scaling dimension of the susceptibility $\Pi^1_{E\lambda\lambda}$ it is possible to extract the scaling behavior of $Q$, as already done in 
Ref.~[\onlinecite{adi_short}], and to take into account the finiteness of the system, as discussed in Section~\ref{scalingO}, without knowing the details of the model. 
The expected scaling behavior is stated in Table \ref{table1}. By applying APT to the specific Ising Hamiltonian under investigation, we could also  extract the numerical 
prefactors of the scaling, as we will explain in the next section.

From the exact solution presented in section \ref{exact}, we evaluate the total  excess energy $Q$ at the final critical point in the scaling limit as:
\be\label{heat}
Q=\sum_{k>0} E_k^{0}\, p^{LZ}_{\rm ex}(k,v)=\sum_{k>0} 4 k \,p^{LZ}_{\rm ex}(k,v).
\ee
In the limit $v L^2 \gg 1$, we convert the sum into an integral to find
\be\label{int_QLZ}
Q={2L v\over \pi}\int_0^\infty dq\, q \,p^{LZ}_{\rm ex}(q)=0.0265 L v.
\ee
In the limit $v L^2\ll 1$: $p^{LZ}_{\rm ex}(q,v)\approx \frac{v^2}{64 k^4}$ and the total excess energy  becomes:
\be
Q\approx \frac{4 v^2}{64}\sum_{m=0}^\infty   \frac{1}{[\frac{\pi}{L}(2 m+1)]^3}= v^2 L^3{7\zeta(3)\over 128 \pi^3},
\label{Q2}
\ee
where we used anti-periodic boundary conditions for fermions which map to periodic boundary conditions for spins (see Ref ~[\onlinecite{Guim_85}] for details).

\subsubsection{Log-fidelity}

The logarithm of the fidelity $F=-\ln(|\langle \psi(0)|0\rangle|^2)$, in the perturbative regime that we are considering, according to  
Eq.~(\ref{M_expect}), can be approximated as: $F\approx \sum_{n\neq 0} |\alpha_n|^2$. Therefore the scaling of the log-fidelity can be extracted from that of the generalized force corresponding to the identity operator: $\mathcal M_\gamma=-\mathbb{I}$, 
as it is easy to see from  our definition in Eq.~(\ref{gen_force}).
Based on the same reasoning as above for the excess energy, the linear response for this observable vanishes, and the coefficient for the quadratic scaling is:
 \be
\Pi^1_{F\lambda\lambda}=\sum_{n\neq 0}{|\langle 0| \partial_\lambda \mathcal H|n\rangle|^2\over (\mathcal E_n-\mathcal E_0)^4}.
\ee
From the exact solution (see also Ref.~\cite{Rams2011}) we can calculate $F$ using:
\be\label{logF}
F=-\sum_{k>0} \ln(1- p^{LZ}_{\rm ex}(k,v)).
\ee
Then in the limit $v L^2 \gg 1$, transforming the sum to an integral we immediately find $F=0.0276\sqrt{v} L $, while in the opposite limit 
$F\approx \frac{v^2}{64}\left(\frac{L}{\pi}\right)^4\sum_{m=0}^\infty   \frac{1}{(2 m+1)^4}=\frac{1}{6144}v^2 L^4$.

\subsubsection{Transverse Magnetization} 

Finally we study the transverse (excess) magnetization
\be\label{Ex}
M_x=\sum_{j} \langle \sigma_j^x\rangle-\sum_{ i} \langle 0|\sigma_j^x|0\rangle.
\ee
According to the definition in Eq.~(\ref{gen_force}), $M_x$ corresponds to the expectation value of the observable $\mathcal M_\lambda=-\partial_\lambda \mathcal H$, 
i.e., the generalized force with respect to the coupling constant that is quenched in time, which in our case is $g(\tau)=1+\lambda(\tau)$. Therefore, from 
Eq.~(\ref{kubo_it}) we expect the scaling:
\be\label{isingM}
M_x \approx -2 v_\lambda g_{\lambda,\lambda},
\ee
where it should be noted that $v_\lambda=\partial_\tau \lambda_\lambda=-v$. We extract the value of $M_x$ from the exact solution a follows: we evaluate for each momentum 
$k$ the expectation value at the end of the process ($\tau=0$): $\langle \hat{\sigma}^z (k)\rangle_{\tau=0}=-(c_1^2-c_3^2)/( c_1^2+c_3^2)$, and the sum over all momenta:
\be\label{Mx_exact_Im}
M_x=-\sum_k\frac{\Gamma(\frac{k^2}{2 v}+1)^2-\frac{k^2}{2 v} \Gamma(\frac{k^2}{2 v}+{1\over2})^2}{\Gamma(\frac{k^2}{2 v}+1)^2+\frac{k^2}{2v} \Gamma(\frac{k^2}{2 v}+{1\over2})^2}.
\ee
Evaluating this expression in the limit $v L^2 \gg 1$  we find $M_x=0.264\sqrt{v} L$, while in the opposite limit 
$M_x\approx \frac{v}{4}\left(\frac{L}{\pi}\right)^2 2 \sum_{m=0}^\infty   \frac{1}{(2 m+1)^2}=\frac{1}{16}v L^2$.

These results from the exact solution are compared in Table \ref{table1} with the ones from APT. The agreement is very good,  we will comment
on this  in more detail in the following section.

We point out that if instead we would perform a quench changing $J(\tau)=1-\lambda=1+v\tau$, with $g=1$ (as we did with QMC in Ref.~\onlinecite{adi_short}), the 
correspondent generalized force $\mathcal M_\lambda$ is now found with $\lambda=-J$. It corresponds to the observable:
\be\label{Ez}
E_z=-J\left[\sum_{\langle ij\rangle} \langle \sigma_z^i\sigma_z^j\rangle-\sum_{\langle ij\rangle} \langle 0|\sigma_z^i\sigma_z^j|0\rangle\right],
\ee
which is the excess interaction energy, or $z$-energy. Therefore, the observables $E_z$ and $M_x$ have the same scaling behavior respectively for a quench of the 
coupling $J$ and of the coupling $g$. 
When we  deal  with the QMC simulations and the numerical solution of the differential equations, it is in practice more convenient
to perform a quench of the $J$ coupling, therefore in the following, when presenting the data  we 
will use the observable $E_z$.

\subsection{Adiabatic perturbation theory for the transverse field Ising model}
\label{adiaPT}

The exact solution for the quench dynamics in imaginary time of the 1D transverse-field Ising model provides a good opportunity to test the APT method
presented in Sec.~\ref{Kubo}. The APT analysis in this case is very accurate, agreeing very well with the exact solution, as shown in Table~\ref{table1}.
The basic ingredient of the APT  is the transition amplitude $\alpha_n(\tau_f=0)$, which in terms of the tuning parameter $\lambda=- v \tau$ becomes~\cite{adi_short}:
\be
\alpha_n(0)\approx \int\limits_0^{\infty} d\lambda \, \langle n|\partial_{\lambda}|0\rangle \exp\left[-\int_0^{\lambda} 
\frac{d\lambda'}{v} (\mathcal E_n(\lambda')-\mathcal E_m(\lambda')\right].
\label{eq_central_lambda}
\ee
For the transverse-field Ising model the lowest excitations correspond to flipping the effective spin from $|-\rangle_k$ to $|+\rangle_k$ (corresponding to exciting 
two Bogoliubov's fermions with opposite momenta) characterized by the matrix element in Eq.~(\ref{mat_elem}) and the energy difference $2\varepsilon_k$. 
Therefore the transition amplitude from the ground to the excited state is:
\beq
&\alpha_k(0)&\approx\frac{1}{2}\int\limits _0^\infty d\lambda {\sin(k)\over \lambda^2+2(\lambda+1)(1-\cos(k))}\nonumber\\
&&\times\exp^{-{4\over v}\int_0^\lambda d\lambda' \sqrt{\lambda^2+2(\lambda+1)(1-\cos(k))}}.
\eeq
This expression simplifies in the slow limit, where we can use the linearized spectrum:
\be
\alpha_k(0)\approx\frac{1}{2}\int\limits _0^\infty d\lambda {k\over \lambda^2+k^2}\exp\left[-{4\over v}\int_0^\lambda d\lambda' \sqrt{k^2+\lambda'^2}\right].
\ee
Using these transition amplitudes instead of the exact expressions found in the previous section we recover the last column in Table~\ref{table1}.

\begin{figure}[htbp]
\begin{center}
\includegraphics[width=8.5cm]{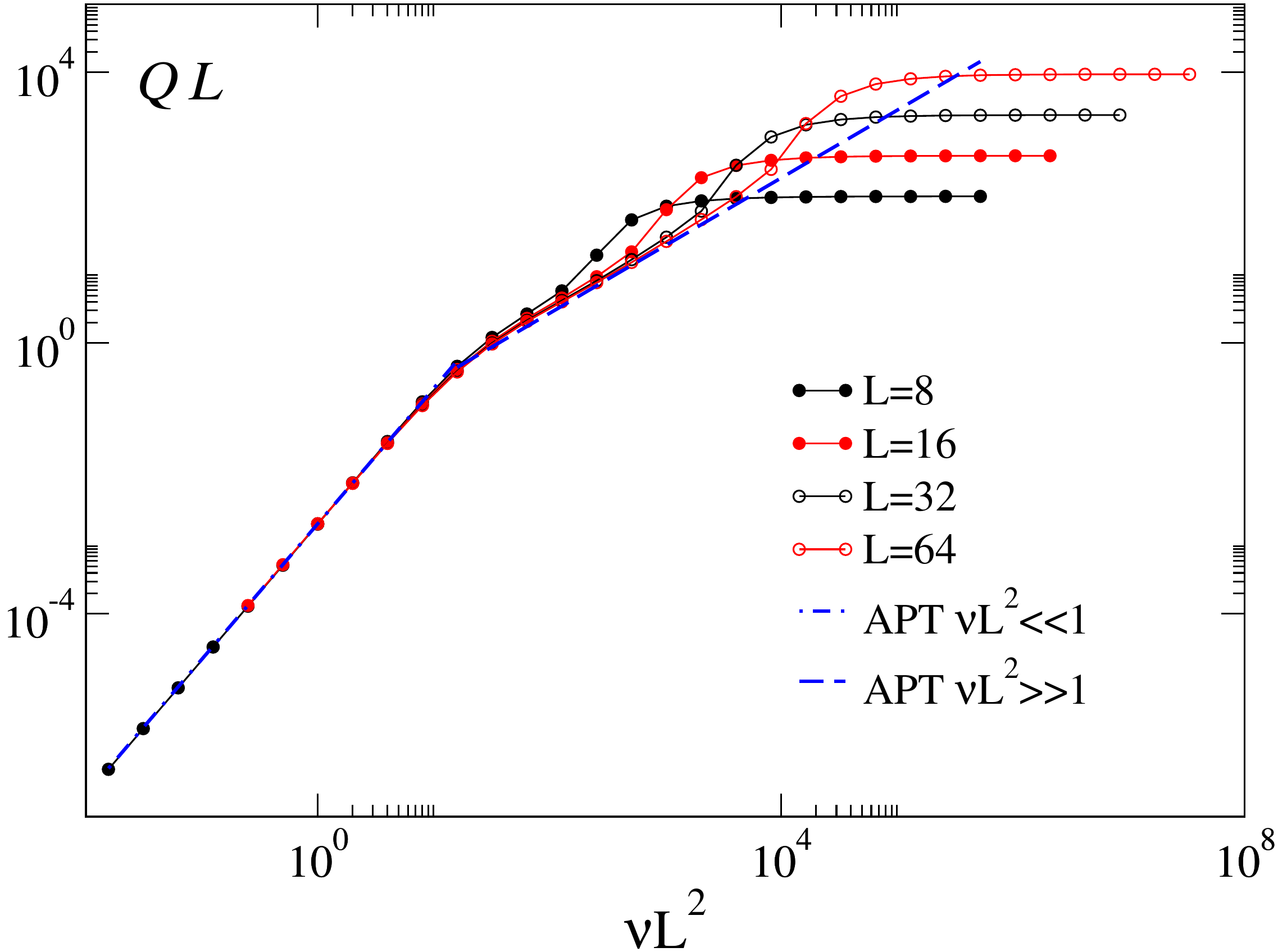}
\caption{(Color online) Results from the exact solution of the  Schr\"odinger equation. The excess energy $Q$ following an imaginary time quench
for different system sizes $L=8,16,32,64$ is shown as a function of $v L^2$, with $v$ varying between $128$ and $10^{-4}$). The data collapse is evident, 
the splitting of the curves for large $vL^2$ is due to finite size effects. The points overlap well with the predictions from APT based on the linearized spectrum,
shown with dashed and dot-dashed straight lines for the high- and low-velocity regimes.  The fitted lines have slope $2$ for $v L^2\ll 1$ and $1$ for $v L^2\gg1$,
in agreement with the APT scaling in Table~\ref{table1}.}
\label{plotQ}
\end{center}
\end{figure}

In Fig.~\ref{plotQ} we plot the excess energy $Q$ for different system sizes as a function of $v L^2$ computed from  the exact solutions of the 
Schr\"odinger equation [solving numerically Eqs.~(\ref{lz1}) and (\ref{lz2})]. In the same plot we also show the APT results for the infinite size. The agreement
between the two methods is excellent when sufficiently large systems are used for the exact solution.

In most situations the exact solution for a time-dependent problem is not readily available. The good agreement we found here suggests that one can make many 
qualitative and even quantitative statements about the dynamics using the adiabatic perturbation theory, which only requires the integration of static quantities. 

\section{Observables in real and imaginary time quenches of the Ising model}
\label{imVSreal}

According to the results presented in the previous sections, it follows that the imaginary-time dynamical protocols give very similar results as the real-time protocols 
considered earlier~\cite{degrandi_09}. This implies, in particular, that imaginary-time quantum evolution with a time evolving Hamiltonian can be used for simulations 
of real-time non-equilibrium dynamics, including, e.g., realizations of the quantum Kibble-Zurek mechanism~\cite{zurek_05}. However, there are still also important 
differences between the two types of dynamics. Firstly, the imaginary-time evolution clearly breaks time-reversal symmetry and this introduces a strong asymmetry 
between the initial and final times of the evolution. Thus, in real-time  dynamics, the system is constantly excited during the evolution and these excitations 
propagate in time. On the contrary, during the imaginary-time evolution the system always relaxes toward the ground state, and the effects of non-adiabaticity 
are visible only when approaching the final state, i.e., the critical point. For example, in real-time evolution the ground state fidelity and the diagonal entropy \cite{Polkovnikov_diagEnt}
(which are observable independent measures of non-adiabaticity) are identical for the time-reversed protocols. In particular, the degree of non-adiabaticity is the same 
if one considers protocols which start or end at the quantum critical point. In imaginary-time evolution this is not the case. If one passes a singularity, like a critical 
point, in a real-time process, then it will always result in non-analyticities in various observables (the defect density in the case of Kibble-Zurek mechanism is an 
example of this). In imaginary-time evolution the singularities in the observables will show up only if one ends the process at this singularity or in its close vicinity. 

In this section we analyze closely the behavior of the observables in the case of quenches of the transverse-field Ising model, comparing the exact solutions for 
the real- and imaginary-time cases. For simplicity we consider here the protocol already analyzed in Ref.~[\onlinecite{adi_short}], where we fix $g=1$ and ramp $J$
linearly in time to end at the critical point, i.e., $J=1+v\tau$ and $J=1+v t$, in imaginary and real time, respectively, with the final time: $\tau_f=t_f=0$. Then 
the scalings of the excess energy, fidelity and magnetization in real and imaginary times look nearly identical. Since real-time evolution in this model 
was analyzed earlier in different papers~\cite{dziarmaga_05, degrandi_09} we will omit the details of the calculation and only present the final results. We note 
that in Ref.~[\onlinecite{degrandi_09}] we analyzed a linear quench where one \emph{starts} at the quantum critical point. Because of the symmetry of the transition 
probabilities with respect to time reversal the analysis applies as well to the process we are interested in here, where one \emph{ends} the quench at the 
quantum-critical point. The only subtlety appears in the analysis of the $x$-magnetization, which is an off-diagonal observable and which depends on the phase 
of the transition amplitude. We will comment on this subtlety below.

\begin{figure}[htbp]
\begin{center}
\includegraphics[width=4.2cm]{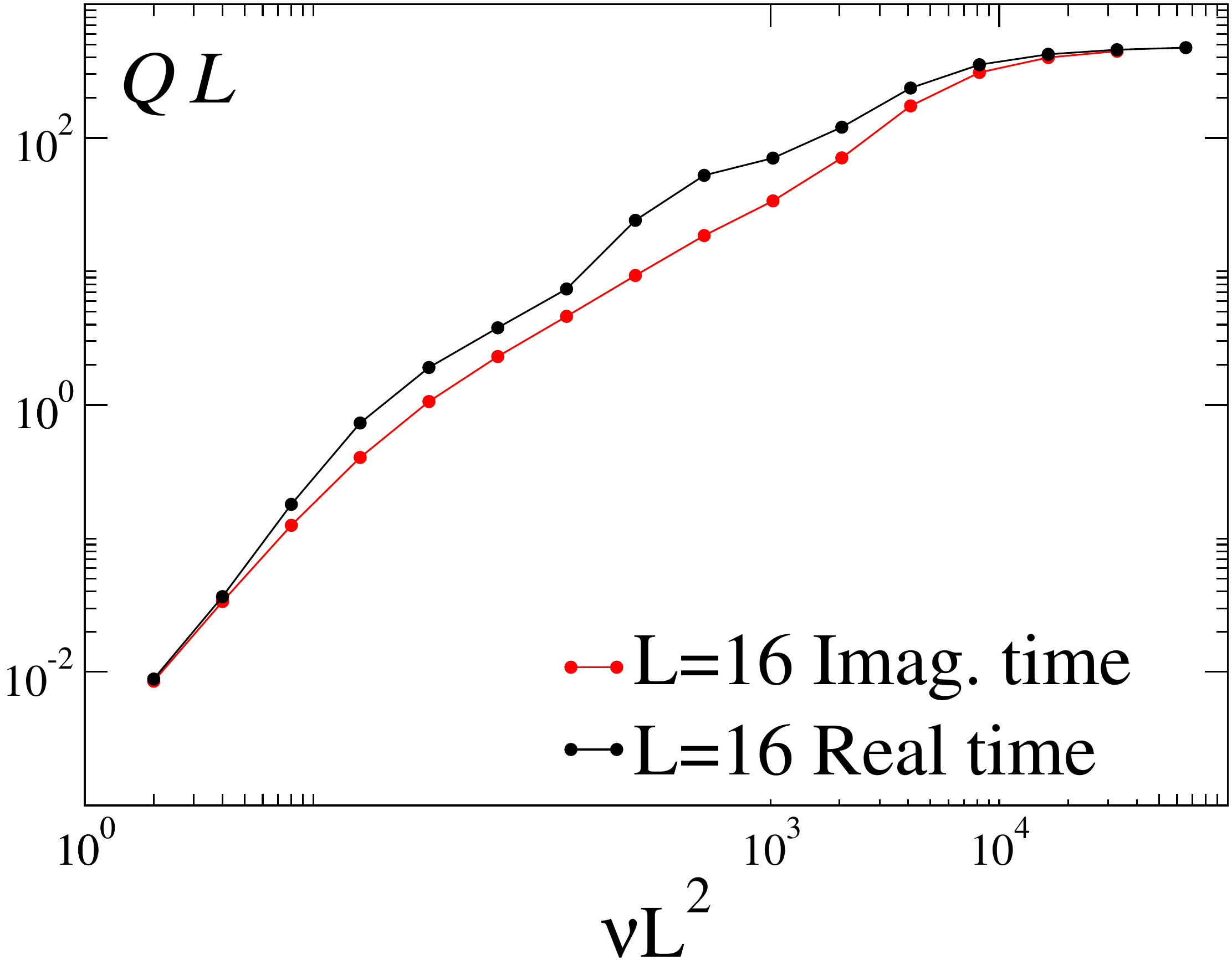}
\includegraphics[width=4.2cm]{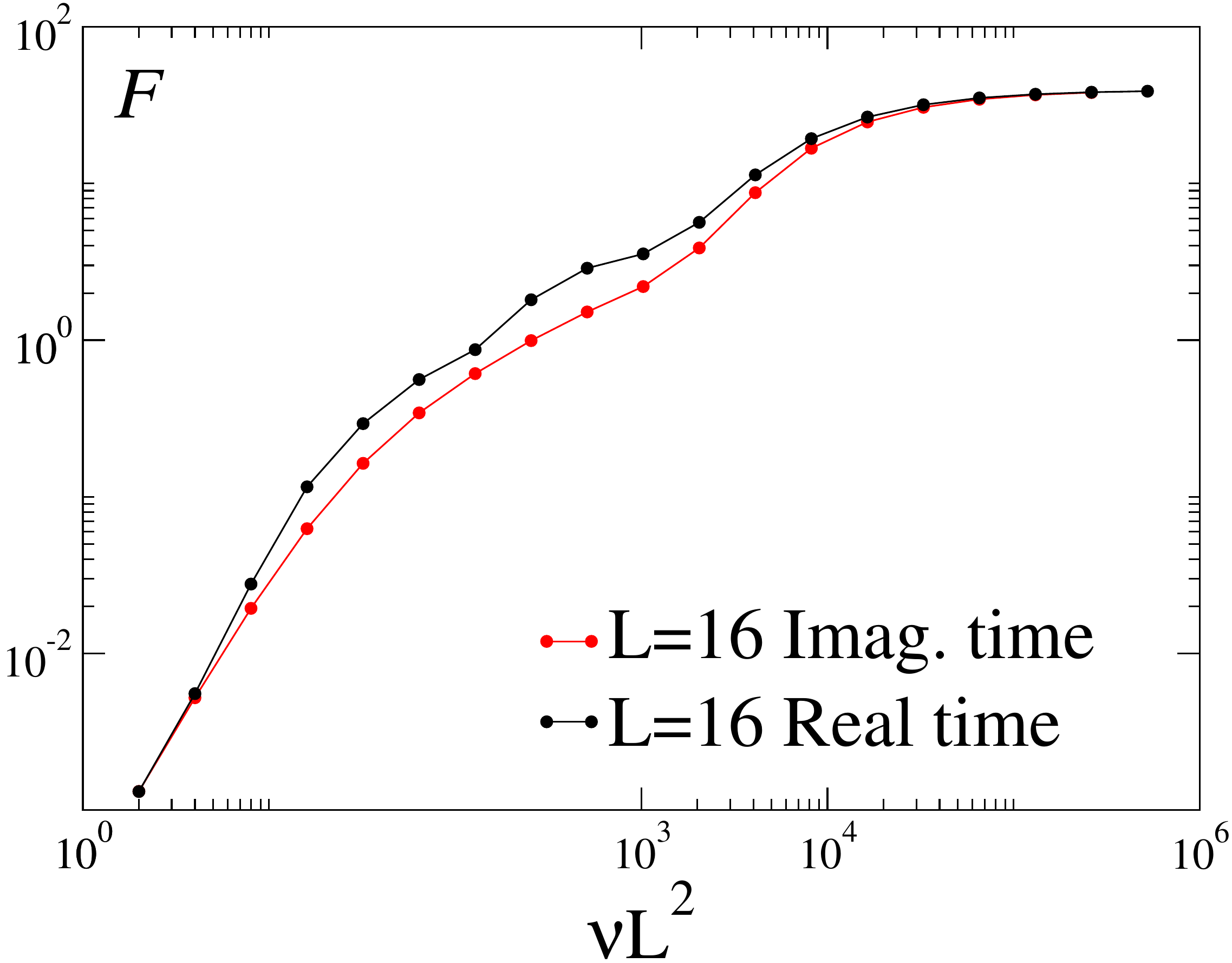}
\includegraphics[width=4.2cm]{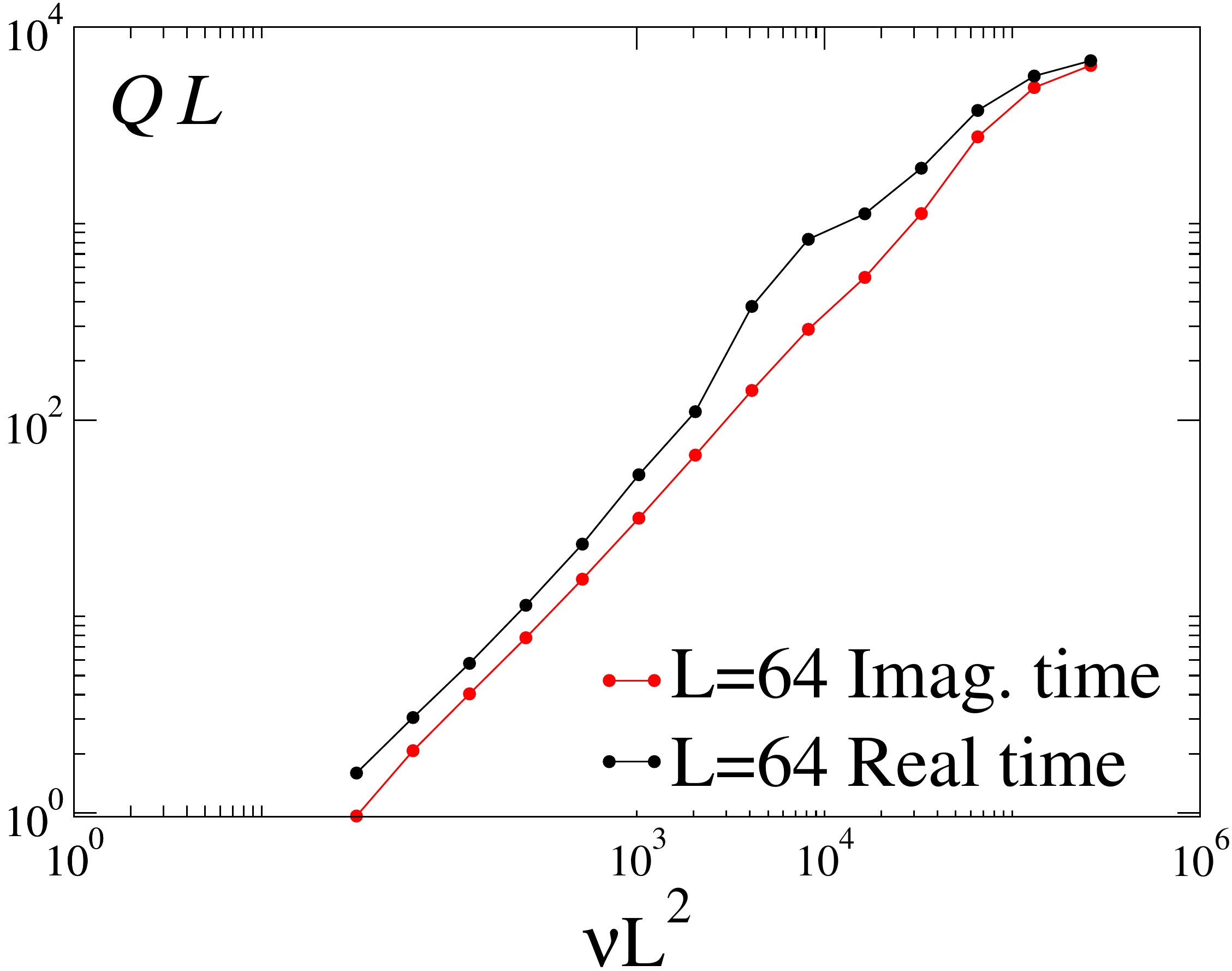}
\includegraphics[width=4.2cm]{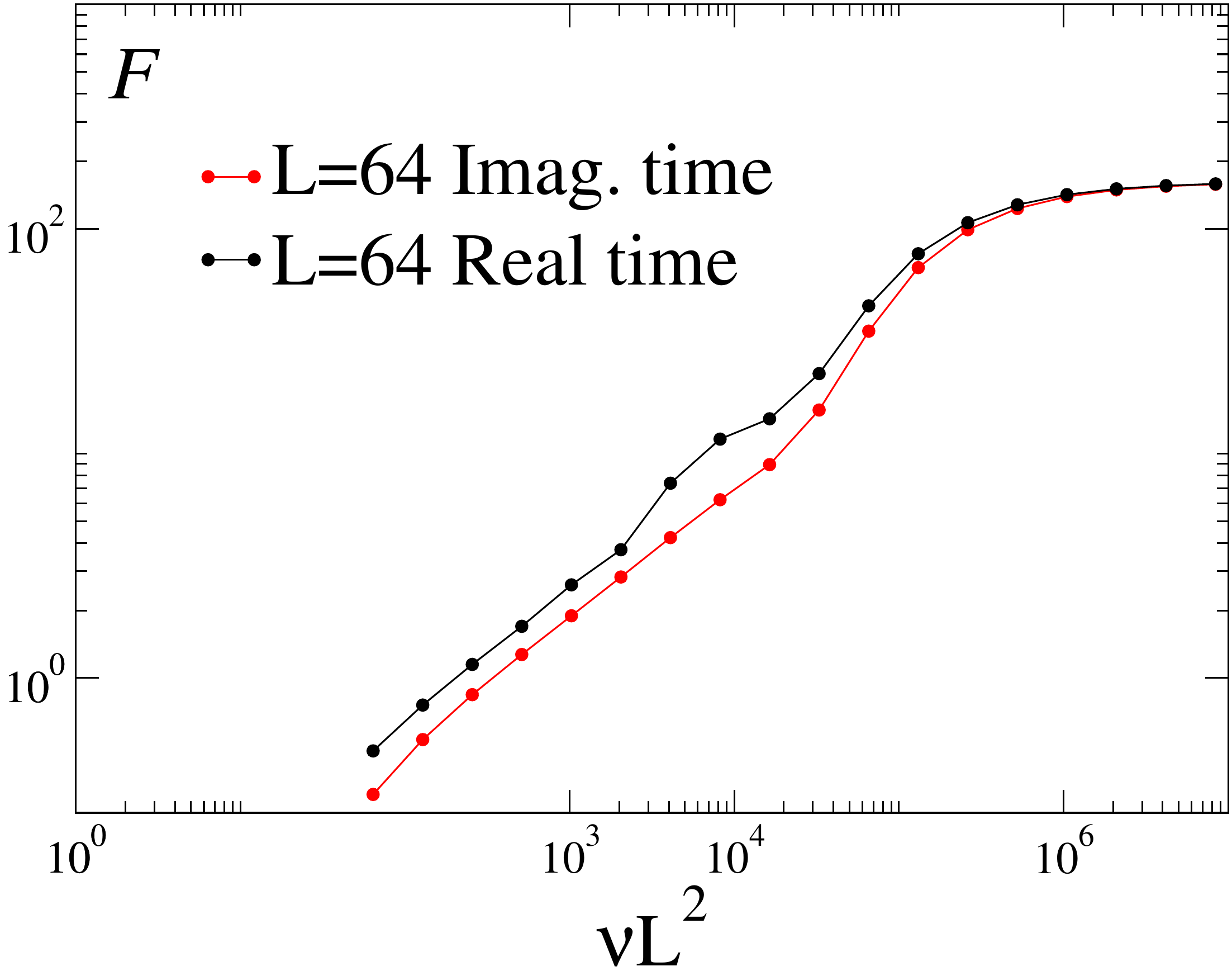}
\caption{(Color online) Comparison of real- and imaginary-time dynamic scaling of the excess energy $Q L$ (left) and the log-fidelity $F$ (right) for system 
size $L=16$ (top) and $L=64$ (bottom).}
\label{plotQReImL}
\end{center}
\end{figure}

\begin{figure}[htbp]
\begin{center}
\includegraphics[width=4.2cm]{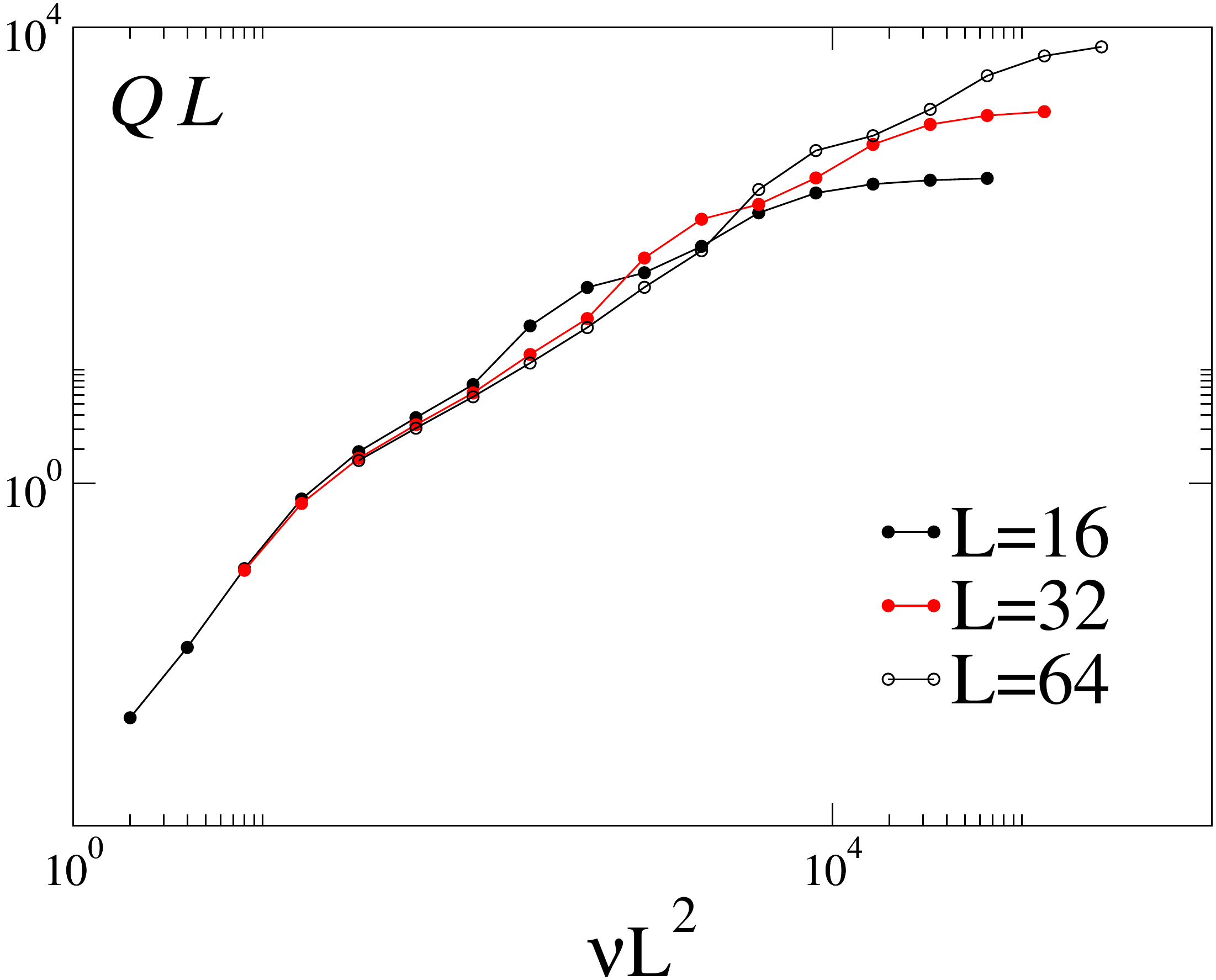}
\includegraphics[width=4.2cm]{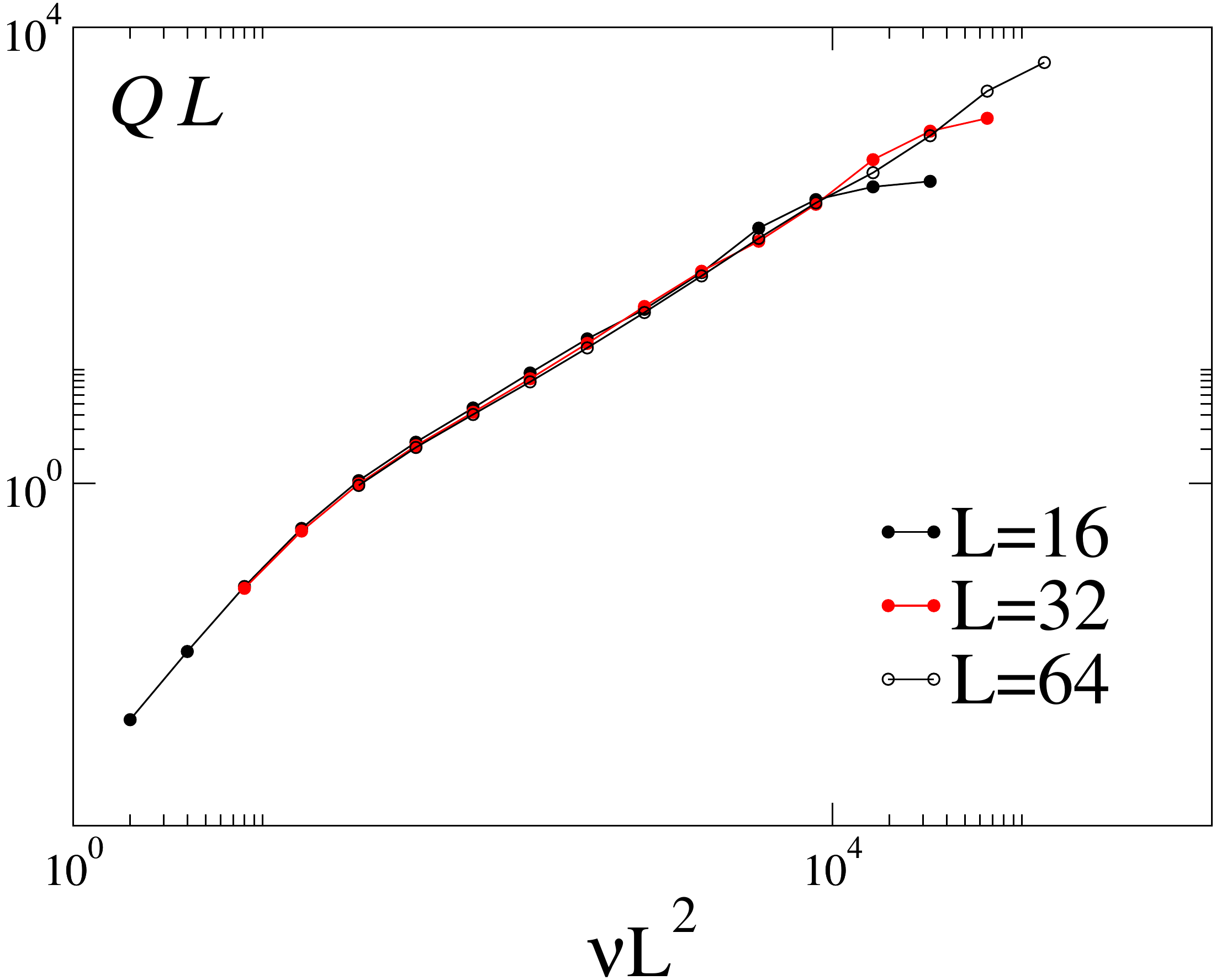}
\includegraphics[width=4.2cm]{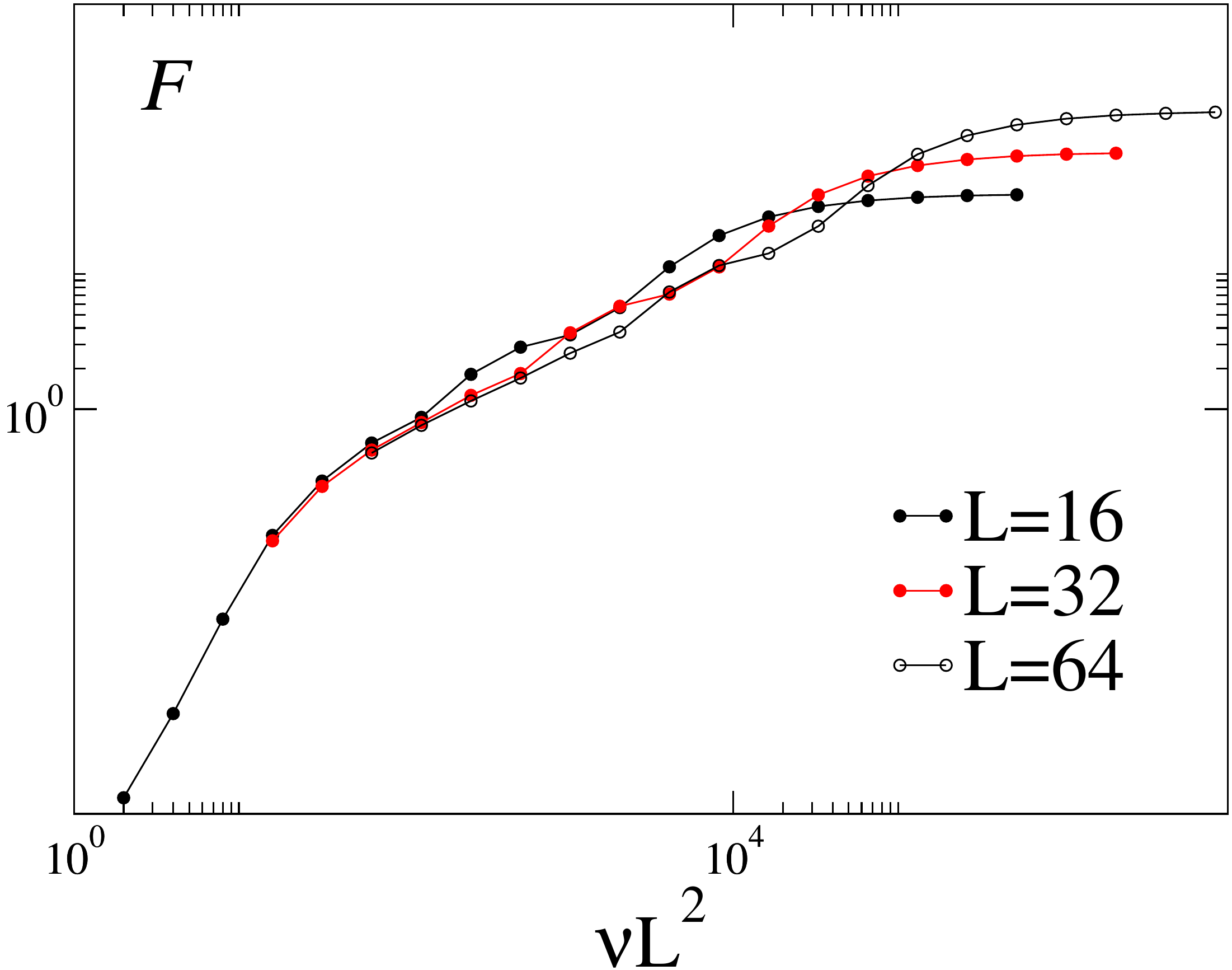}
\includegraphics[width=4.2cm]{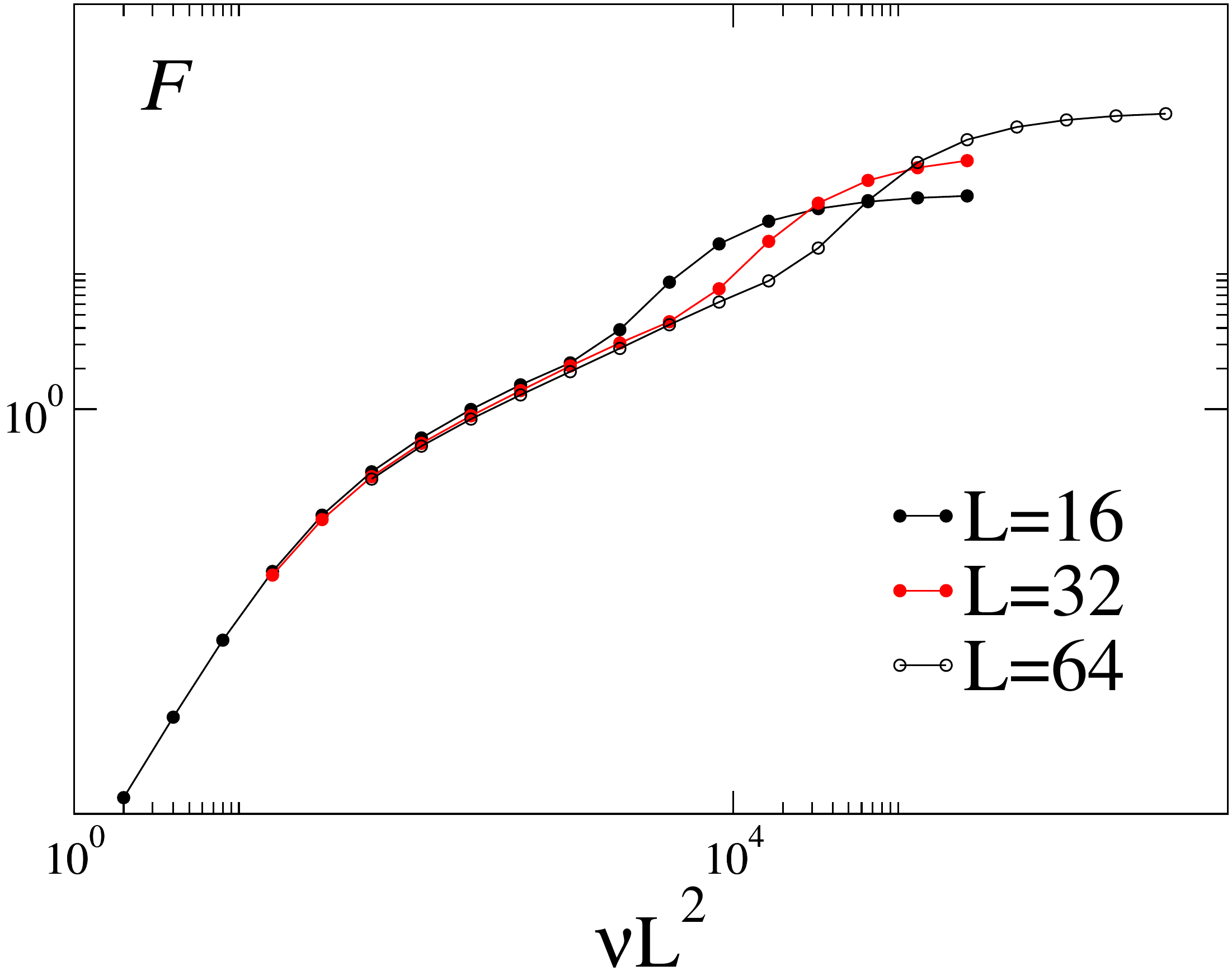}
\caption{(Color online) Data-collapse plot for the excess energy $Q$ (top) and the log fidelity $F$ (bottom) based on real- (left)  and imaginary-time (right) 
dynamics for different system sizes. In the regime of large $vL^2$ the splitting of the curves is due to finite-size effects. The real-time case show more 
oscillating behavior than the imaginary-time case.}
\label{plotQReIm}
\end{center}
\end{figure}

\begin{figure}[htbp]
\includegraphics[width=.42\textwidth]{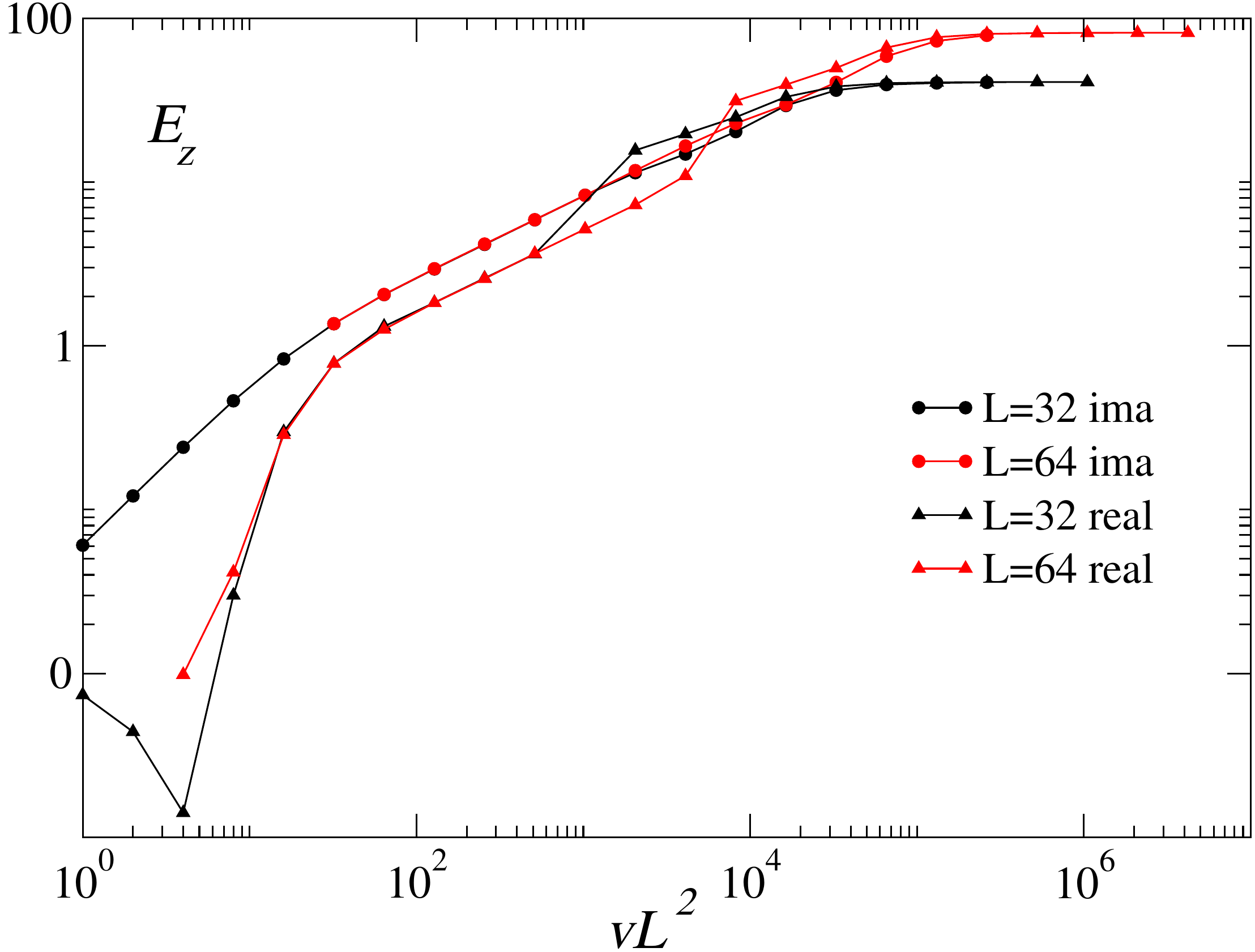}
\caption{(Color online) The  excess interaction energy $E_z$ (which has the same scaling as the magnetization $M_x$) in real- and imaginary-time dynamics for 
system sizes $L=32$ (black) and $L=64$ (red). For $vL^2\gg 1$ the slopes are the same, except for a shift due to a different prefactor. The splitting of the curves  for large $vL^2$ is due to finite-size effects. For $vL^2\ll 1$ the real-time case changes drastically, decaying to zero exponentially.}
\label{Ez_ReIm}
\end{figure}

In Table~\ref{table2} we present the comparison of the scaling of several observables for linear quenches to the QCP in real and imaginary time, obtained from the 
exact solution of the transverse-field Ising model (see Sec. \ref{exact} for the imaginary-time case and Refs.~\cite{dziarmaga_05, degrandi_09} for the real-time 
case). In Figures~\ref{plotQReImL},~\ref{plotQReIm}, and \ref{Ez_ReIm} we plot the corresponding quantities obtained by solving numerically the  Schr\"odinger equation.
The definition of the observables was given in the previous Section~\ref{Ising}.  As mentioned before,  since we are quenching $J$ (and not $g$, to be consistent with 
the protocol used in QMC simulation presented in following sections), the observable which gives the fidelity susceptibility in the linear response is the excess interaction energy $E_z$ [see Eq.~(\ref{Ez})]. It is expected to scale in the same way as $M_x$.

Overall we find very good agreement and almost identical behavior between the imaginary and the real time cases. A more careful analysis is nevertheless necessary.
For the diagonal observables, the excess energy $Q$ and the log-fidelity $F$, the scaling behaviors are the same in real and imaginary time and in agreement 
with the APT predictions presented in the previous section. In particular, in the limit $vL^2\ll 1$ even the prefactors coincide---indeed, in this limit the analytic 
expression are identical. In the opposite regime $vL^2\gg 1$ the prefactors are slightly different. In this limit, the real-time dynamics presents a more  oscillating 
behavior: see for instance the plots of the  excess energy in Fig.~\ref{plotQReIm}. A similar behavior was also observed in 
Refs.~\cite{kolodrubetz_11,kolodrubetz_11a}.

The case of the excess $x$-energy or magnetization along the $x$-direction [as defined in Eq.~(\ref{Ex})] requires more attention. Indeed this quantity, as mentioned 
before, corresponds to the generalized force with respect to the coupling $\lambda$ that drives the dynamics. Working out the asymptotic scaling behavior from the 
scaling dimension in the limit of $vL^2\gg 1$  we find $M_x\sim  \sqrt{v} L$ in both real and imaginary times, according to Eqs.~(\ref{kubo_rt}) and ~(\ref{kubo_it}). 
Concerning the limit $vL^2\ll 1$, in imaginary time, from the exact solution [see Eq.~(\ref{Mx_exact_Im})] we know that $M_x\sim v L^2$. In the real-time case, 
from analyzing the exact solution we can infer that the behavior for small $vL^2$ is non analytic, decaying exponentially as $\sim e^{-{\pi^3\over vL^2}}$. Such behavior
is visible in the plot  in Fig.~\ref{Ez_ReIm}; for large values of $vL^2$ (but not too large, as finite-size effects also are apparent) the slopes of the real- and 
imaginary-time functions are the same, the data being shifted by a factor of $2$ according to the predictions. For $vL^2 \ll 1$ the imaginary-time function decays 
analytically with slope $1$ as expected, while in the real-time case there is a more rapid drop reflecting the non-analyticity of the function.

\begin{table}[t]
\begin{center}
\begin{tabular}{|c|c|c|c|c|}
\hline 
 & \multicolumn{2}{|c|}{$vL^2\ll 1$}    & \multicolumn{2}{|c|}{$vL^2\gg 1$}  \tabularnewline
\hline
 Observable & Real &  Imag. & Real  &  Imag. \tabularnewline
 \hline
  $M_x$ &  0 & ${1\over 16} v L^2$ 
& $0.16\sqrt{v}L$  & $0.26\sqrt{v}L$
\tabularnewline
\hline
$Q$ & \multicolumn{2}{|c|}{${7\zeta(3)\over 128 \pi^3} v^2 L^3$} &    $0.038\, vL$ & $0.0265\, vL$\tabularnewline
\hline
$F$ & \multicolumn{2}{|c|}{${1\over 6144} v^2 L^4 $} &  $0.035 \sqrt{v} L$  & $0.0276 \sqrt{v} L$\vspace{0.05cm}\tabularnewline
\hline
\end{tabular}
\caption{Results from the exact solution of the 1D transverse-field Ising model: scaling forms for the magnetization, the excess energy, and log-fidelity 
with the quench velocity $v$ and the system size $L$ in real and imaginary time for different regimes.}
\label{table2}
\end{center}
\end{table}

\section{Application: detection of quantum-critical points}
\label{detectQCP}

\begin{figure}[htbp]
\begin{center}
\includegraphics[width=8cm]{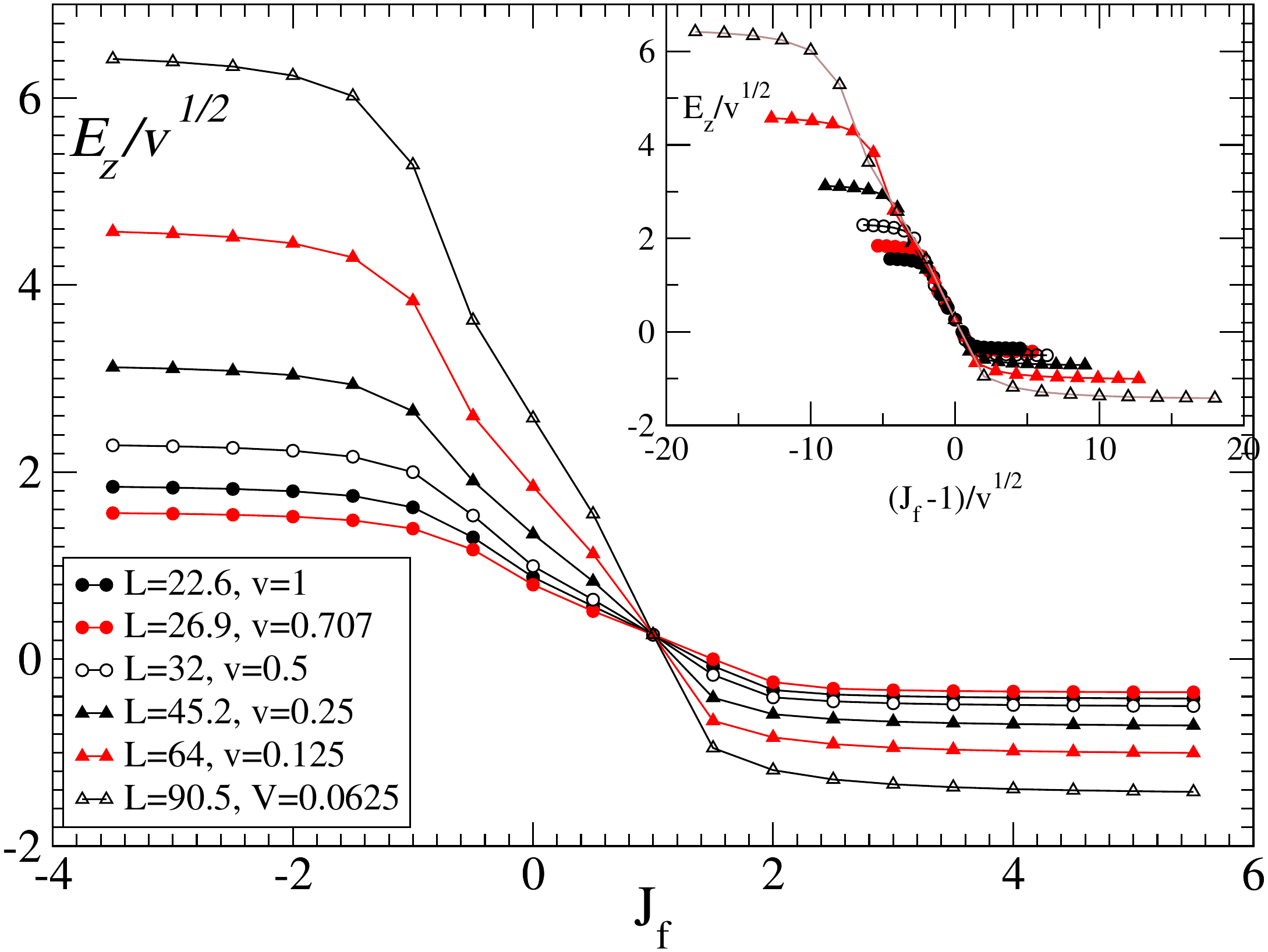}
\includegraphics[width=8cm]{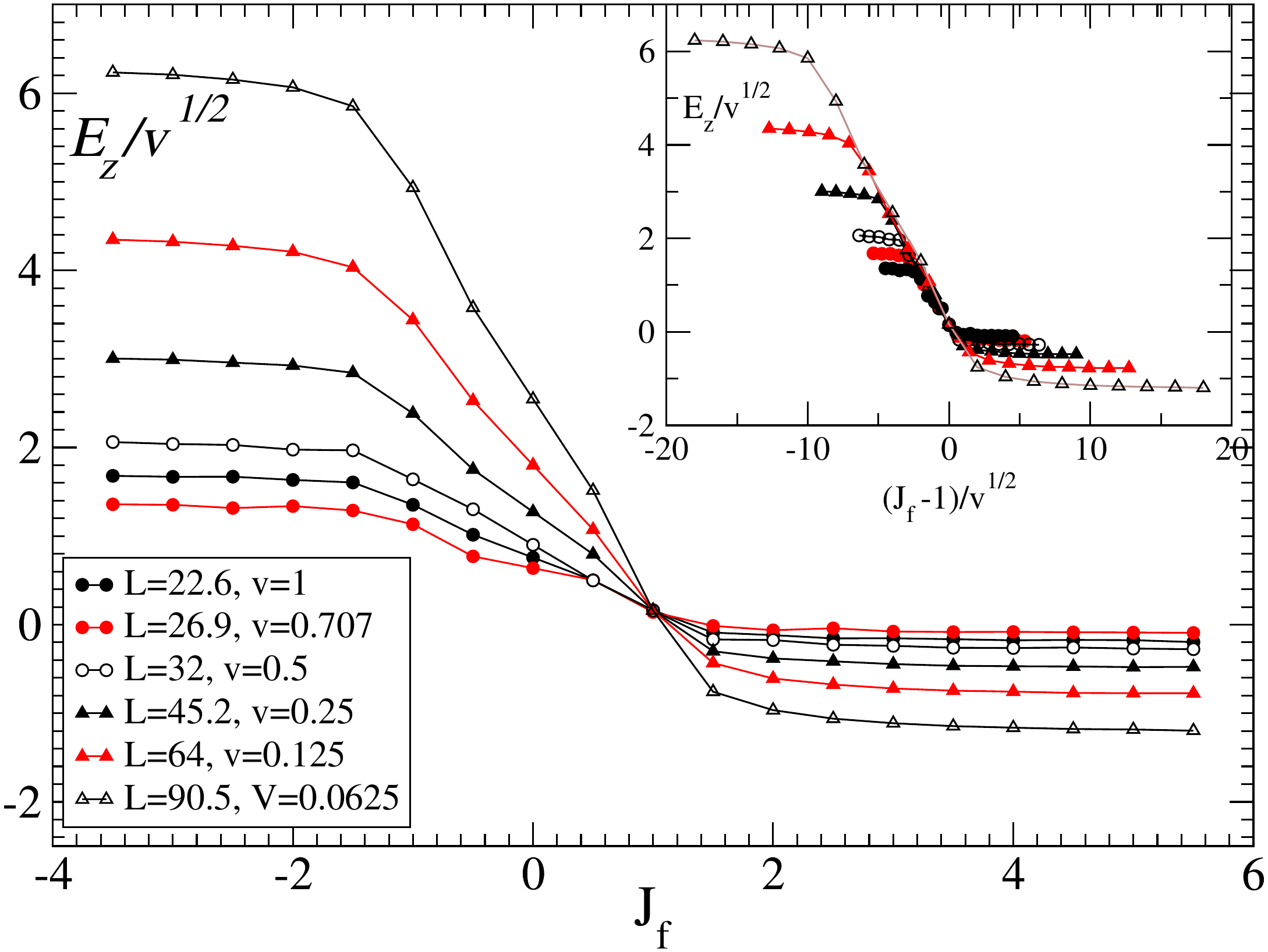}
\caption{(Color online) Real-time (bottom graph) and imaginary-time (top panel) results for quenches ending at variable finite amplitude $J_f$. 
The rescaled $z$-energy $E_z/\sqrt{v}$ is shown for different quench velocity $v$  and system size $L$, with the product $vL^2$ fixed at $512$.
In both cases the curves cross around the location of the quantum critical point $J=1$. The inset shows the same data with the $x$-axis rescaled appropriately 
to achieve the data collapse.}
\label{plotQCPvL}
\end{center}
\end{figure}

\begin{figure}[htbp]
\begin{center}
\includegraphics[width=8cm]{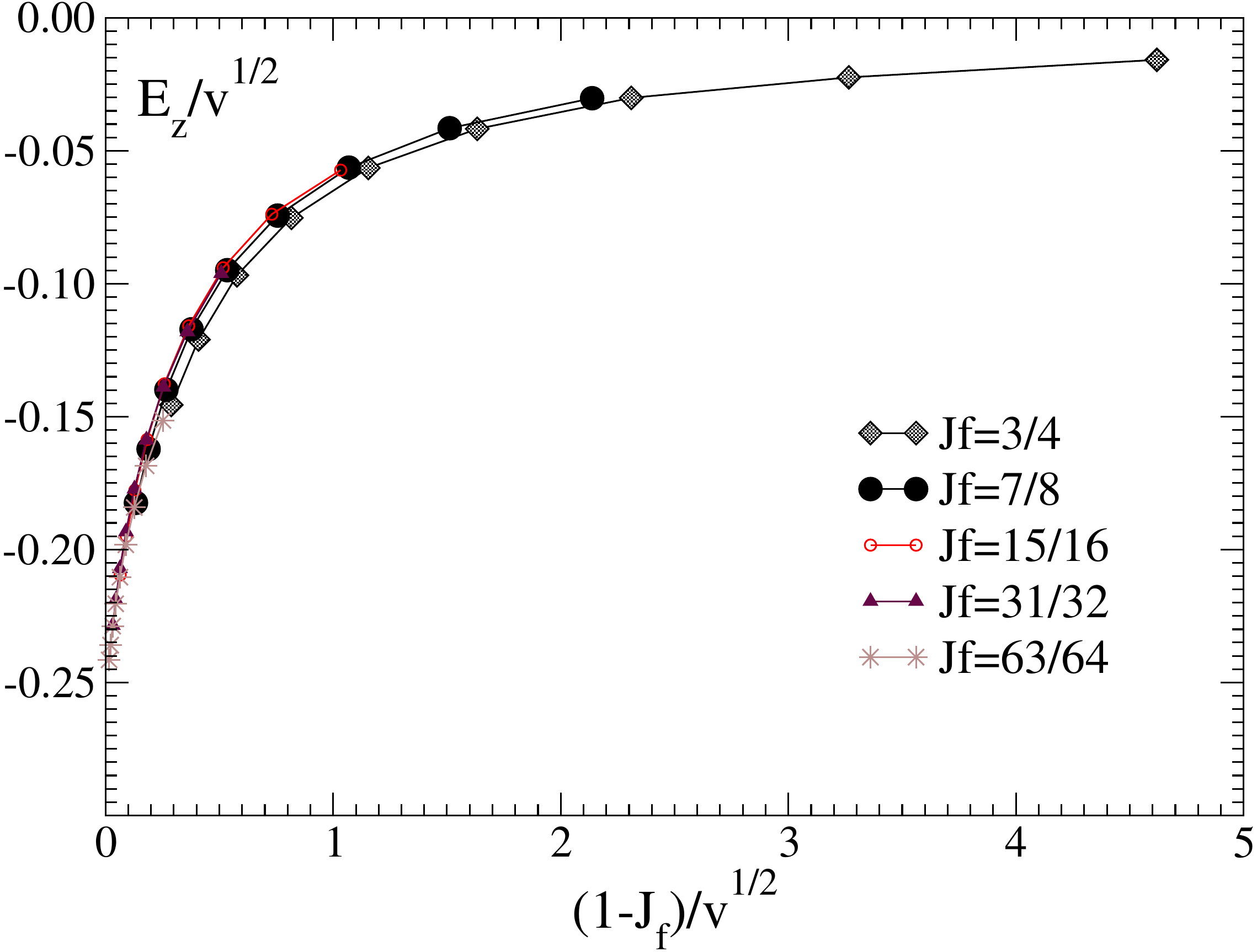}
\caption{QMC results:  data collapse for the $z$-energy $E_z$ for a linear quench of the coupling $J$ ending at different values
$J_f$ just before the QCP.}
\label{plotQMC_Ez}
\end{center}
\end{figure}

A useful application of the universal scaling presented in the previous sections can be found by considering quenches in either real or imaginary time that 
sweep through the QCP and end at different final amplitudes $\lambda_f\neq \lambda_c$.  In this case, as discussed in Sec.~\ref{scalingO}, the length scale 
$\xi_\lambda(\vec\lambda)\sim|\vec\lambda-\vec\lambda_c|^{-\nu}$ is not diverging anymore and participates in the scaling behavior along with 
$\xi_v\sim |\vec v|^{-\nu\over z\nu+1}$ and the system size $L$. For a generalized force $M_\gamma$ we then expect a scaling behavior as the one 
given in Eq.~(\ref{scaling_neq}):
\beq
M_\gamma(\vec \lambda,\vec v)&=&{\rm const}+L^{-\mu_\gamma} f_\gamma(L/\xi_\lambda(\vec\lambda), L/\xi_v)\nonumber\\
&=&{\rm const}+L^d v^{(d+\mu_\gamma)\nu\over 1+\nu z } \tilde{f}_\gamma\left( {|\lambda_f-\lambda_c|\over v^{1/(z\nu +1)}},v L^{z+1/\nu}\right),\nonumber\\
\label{scaling2}
\eeq
as was already suggested in Ref.~\cite{adi_short}.  If we perform several quenches changing the final amplitude $\lambda_f$ and plot for each of them 
the rescaled quantity $M_\gamma/v^{(d+\mu_\gamma)\nu\over 1+\nu z }$, we expect all the curves (asymptotically for sufficiently large $L$) to cross at the location 
of the QCP, since when $\lambda_f=\lambda_c$ the rescaled quantity does not depend on $v$ anymore. We have performed such an analysis for an imaginary-time quench 
of the form $J(\tau)=1-\lambda=1+v\tau$, and the correspondent real-time one (replacing $\tau$ with $t$). Sweeping across the critical point, i.e., starting from a 
negative $\lambda$ and ending at some positive value, and looking at the $z$-energy $E_z$ as defined above in Eq.~(\ref{Ez}), we expect:
\be\label{Ez_uni}
E_z=L \sqrt{v} f_\lambda\left( {|J_f-J_c|\over \sqrt{v}},v L^2\right).
\ee
In Fig.~\ref{plotQCPvL} we show the results for imaginary- (top graph) and real-time (bottom graph) quenches based on the numerical solution of the Schr\"odinger equation. 
The expected behavior is confirmed in both cases, all the lines cross around $J=1$ where the QCP is located. Therefore, through this type of analysis it would be 
possible to locate the position in the parameter space of a QCP of a system that cannot be solved exactly and of which the position of the critical point is not known.
The universal behavior described by Eq.~(\ref{Ez_uni}) is furthermore confirmed by the collapse of the data when plotting versus the rescaled quantity $(1-J_f)\sqrt{v}$:
see the insets of Fig.~\ref{plotQCPvL}. As expected, a similar collapse is also observed if we perform a quench that does not reach the QCP but ends just before it, 
as we show by the data in Fig.~\ref{plotQMC_Ez}. These results were obtained by numerical simulation of the 1D Ising model with the NEQMC method introduced in 
Ref.~\cite{adi_short} (and discussed also above in Sec.~\ref{qmcmethods}), using it to perform a linear quench ending at different values of the 
final amplitude $J_f$ and approaching $J_c=1$.

\section{Universal relaxation to equilibrium after a quench}
\label{relax}

\begin{figure}[ht]
\begin{center}
\includegraphics[width=.45\textwidth]{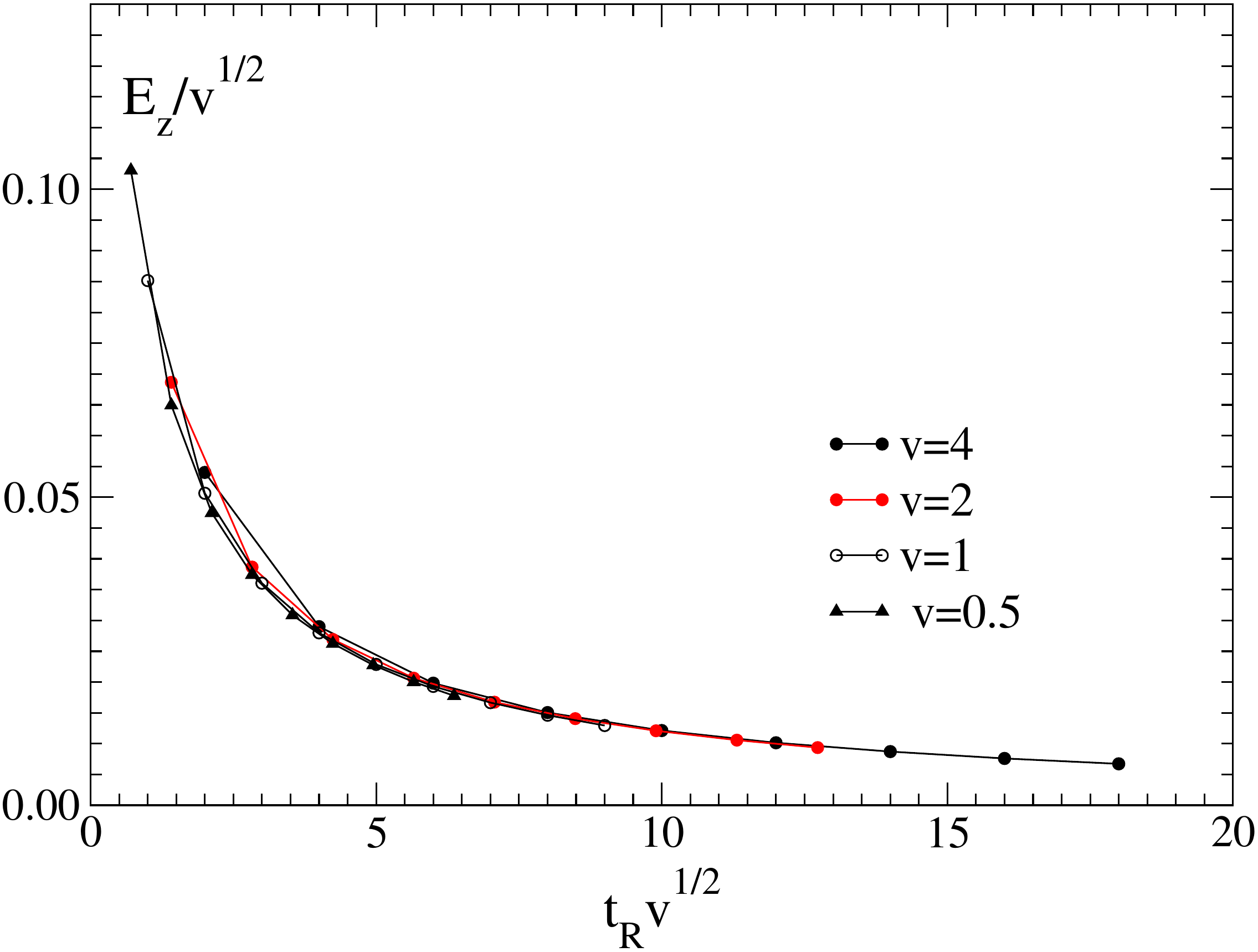}
\caption{Universal relaxation dynamics in imaginary time after a linear quench.}
\label{RelaxIma}
\end{center}
\end{figure}

\begin{figure}[ht]
\begin{center}
\includegraphics[width=.44\textwidth]{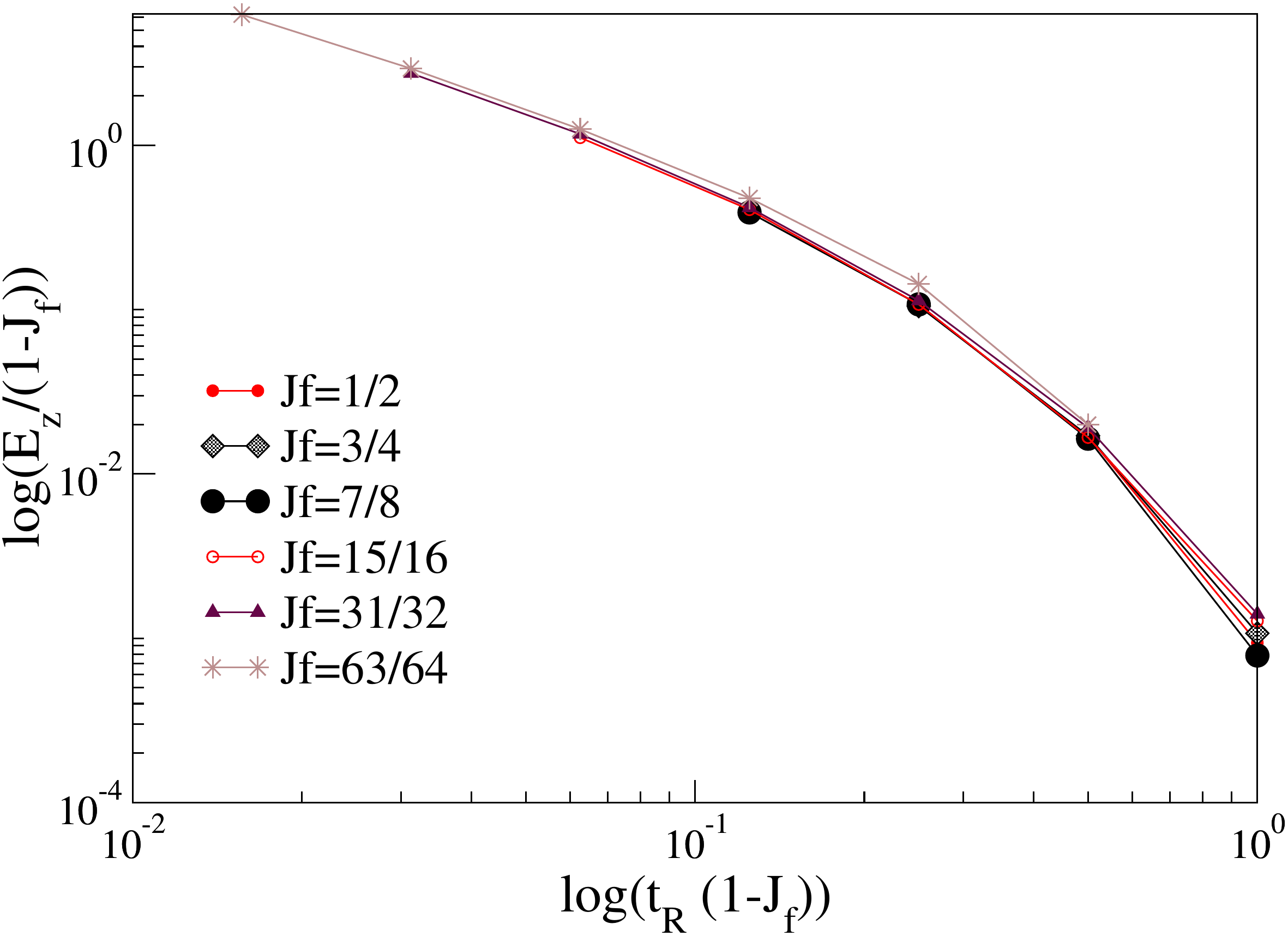}
\caption{QMC data: relaxation dynamics of the interaction energy $E_z$ in imaginary time after a sudden quench.}
\label{Relax_QMC_sudEz}
\end{center}
\end{figure}

Up to this point we have been concerned with the scaling of the observables right \emph{at the end} of a quench, in particular at the final time, when the 
Hamiltonian of the system has reached the QCP $\lambda_f=\lambda_c$ or, as in the previous section, some other amplitude $\lambda_f\neq \lambda_c$. Here we want to address the scaling behavior 
that follows \emph{subsequent to the quench}, when the system starts relaxing governed by a fixed Hamiltonian. 
Therefore we let the system evolve after the end of a quench with the fixed final Hamiltonian  for a variable length of time, that we call $t_R$, and we look at the behavior as a function of $t_R$, that we call the relaxation time.
Based on the scaling arguments that have lead 
us to Eq.~(\ref{scaling_neq}), we argue that, if we let the system evolve after the quench for a time $t_R$, then we expect [for a generic $r$-th 
power quench as in Eq.~(\ref{lambda_rt})]:
\be
M_\gamma(\vec \lambda,\vec v)\sim L^{-\mu_\gamma} f_\gamma\left(t_R v^\frac{\nu z}{z\nu r+1}, L^z/t_R, |\lambda_f-\lambda_c|^{z\nu} t_R\right).
\label{scaling_relax}
\ee
This means that the relaxation time itself comes into play in the universal scaling behavior as an additional ``length'' scale to be compared to the other
characteristic lengths of the system. For sudden quenches this conjecture was recently suggested and tested in Ref.~\cite{Dallatorre2012}. If the system size is large enough, for instance, we expect the quantity $t_R v^\frac{\nu z}{z\nu r+1}$ to be the 
rescaled variable that characterize the universal relaxation after quenches with different velocities. As before, we have checked this behavior for the 
$z$-component excess energy $E_z$ from the exact solutions [solving numerical Eqs.~(\ref{lz1}) and (\ref{lz2})] for a linear quench, see Fig.~\ref{RelaxIma}, 
and with NEQMC for a sudden quench see Fig.~\ref{Relax_QMC_sudEz}.

\section{Summary and conclusions}

We have presented an overview of general aspects of non-adiabatic response of physical observables to slowly changing parameters, both in imaginary (Euclidean) 
and real time. There are many similarities between the imaginary- and real-time response, which we demonstrated by calculating the leading first- and second-order 
non-adiabatic corrections of physical observables. We identified the corresponding susceptibilities and expressed them through the non-equal time correlation 
functions. In particular, we extended the traditional Kubo response theory to describe the response of systems to perturbations which are slow but can be 
arbitrarily large in amplitude. The components of the geometric tensor (the metric tensor and the Berry curvature) naturally emerge as response functions 
of physical observables to the quench velocity.

Both real- and imaginary-time dynamics near continuous phase transitions can be used to analyze universal non-adiabatic response of the system and 
extract static and dynamic critical exponents, using a generalized non-equilibrium scaling theory, which we also further elaborated here. Importantly, 
imaginary-time dynamics is amenable to powerful Monte Carlo simulation methods. We briefly reviewed two different QMC algorithms which directly
implement quantum dynamics for interacting systems. They have the same range of practical applicability (avoidability of sign problems) as conventional
equilibrium finite-temperature or ground-state projection methods. 

We illustrated the utility of the general theoretical formalism using the particular example of the transverse-field Ising model in one dimension. 
Using both exact treatments (through the standard mapping to fermions) and QMC simulations, we found that imaginary- and real-time dynamical responses 
indeed are very similar near the critical point, for all physical observables examined. We also found excellent agreement with predictions of
adiabatic perturbation theory. We illustrated how one can use the non-equilibrium finite size scaling to accurately extract the transition point and 
the critical exponents.

The ideas presented in this article have many potential applications, including (i)  analysing universal dynamical response near quantum phase 
transitions with unknown dynamical exponent, e.g., in disordered systems; (ii) applying QMC methods to implement imaginary-time quantum annealing 
and (using the similarity of non-adiabatic response in real and imaginary times) making predictions concerning real-time quantum annealing  protocols; 
(iii) using non-adiabatic response of physical observables to directly extract the Berry curvature and the metric tensor (including the fidelity 
susceptibility) either experimentally or numerically.

\begin{acknowledgements} 
We would like to thank Cheng-Wei Liu for collaboration on the QAQMC method and for generating the data presented in Fig.~\ref{susqaqmc}.
This work was supported by the NSF under Grant No.~NSF PHY11-25915 (AP and AWS). AP also acknowledges support from Grants
NSF DMR-0907039 and AFOSR FA9550-10-1-0110, and from the Simons and Sloan Foundations. 
\end{acknowledgements}

\bibliography{adi_long}

\end{document}